\documentclass[letterpaper,twocolumn,10pt]{article}
\usepackage{hegroup}

\usepackage[numbers]{natbib}
\usepackage{graphicx}
\usepackage{amssymb}
\usepackage{xspace}
\usepackage{cleveref}
\title{Humanizing LLMs: A Survey of Psychological Measurements with Tools, Datasets, and Human-Agent Applications}

\date{}
\author{
Wenhan Dong\textsuperscript{1}\thanks{Equal contribution.}  \ \ \ 
Yuemeng Zhao\textsuperscript{1}\footnotemark[1]  \ \ \ 
Zhen Sun\textsuperscript{1}  \ \ \ 
Yule Liu\textsuperscript{1}  \ \ \ 
Zifan Peng\textsuperscript{1}  \ \ 
Jingyi Zheng\textsuperscript{1}  \ \ \ \
Zongmin Zhang \textsuperscript{1}  \ \ \\
Ziyi Zhang \textsuperscript{1}  \ \ \ 
Jun Wu \textsuperscript{2} \ \ \
Ruiming Wang \textsuperscript{2} \ \ \
Shengmin Xu\textsuperscript{3} \ \ \
Xinyi Huang\textsuperscript{4} \ \ \
Xinlei He\textsuperscript{1}\thanks{Corresponding author (\href{mailto:xinleihe@hkust-gz.edu.cn}{xinleihe@hkust-gz.edu.cn}).} \ \ \ 
\\
\\
\textsuperscript{1}\textit{Information Hub, Hong Kong University of Science and Technology (Guangzhou)} \ \ \ 
\\
\textsuperscript{2}\textit{School of Psychology, South China Normal University} \ \ \ 
\\
\textsuperscript{3}\textit{College of Computer and Cyber Security, Fujian Normal University} \ \ \ 
\\
\textsuperscript{4}\textit{College of Cyber Security, Jinan University} \ \ \ 
\\
}

\newcommand{\mypara}[1]{\smallskip\noindent{\bf {#1}.}\xspace}
\usepackage{cuted}

\begin{document}
\maketitle

\begin{strip}
\vspace{-7em}
\centering
\includegraphics[width=\textwidth]{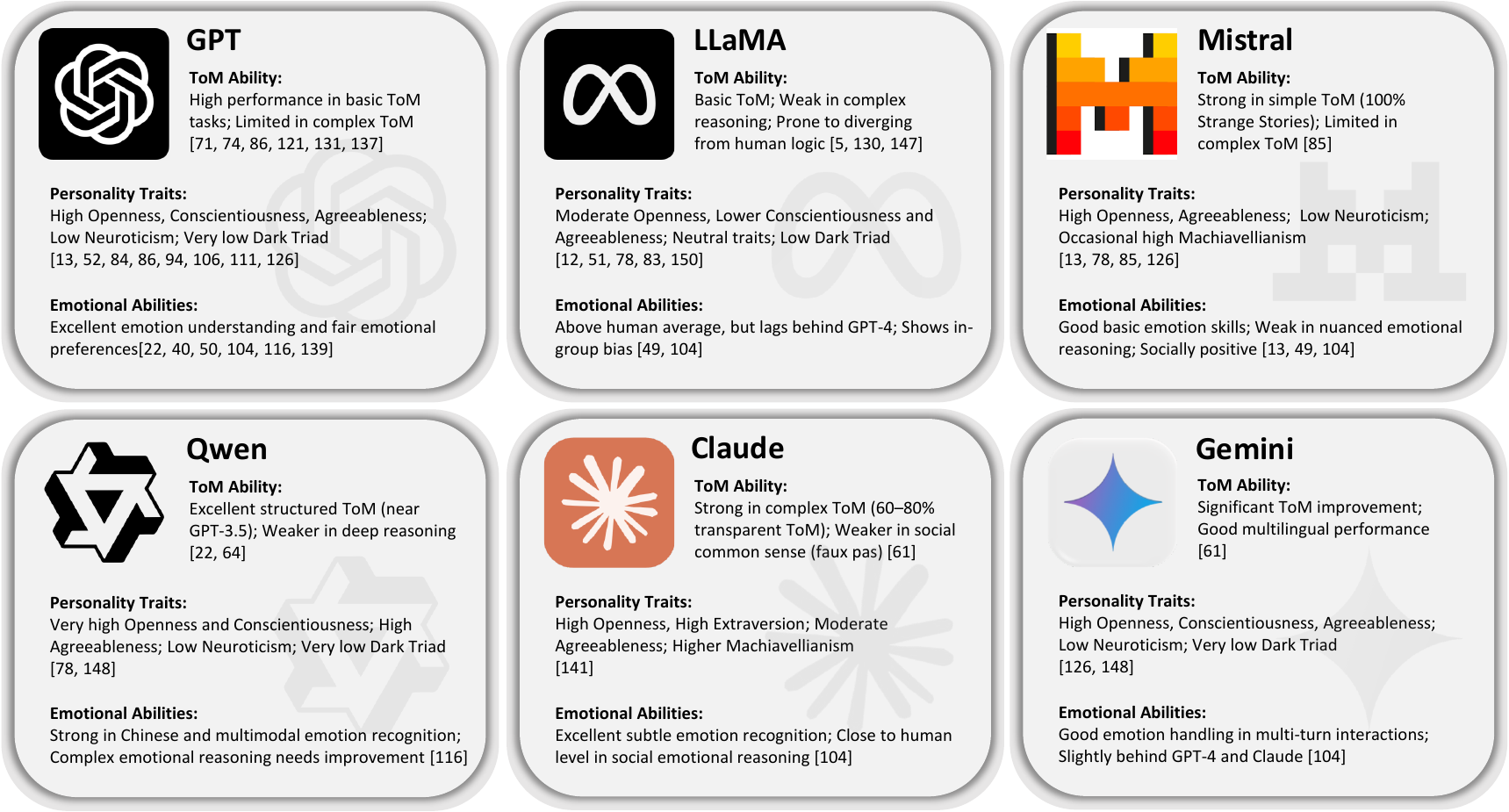}
\captionof{figure}{Psychological ID of LLMs.}
\label{LLM_Traits}
\end{strip}

\begin{figure*}[htbp]
  \centering
  \includegraphics[width=0.8\textwidth]{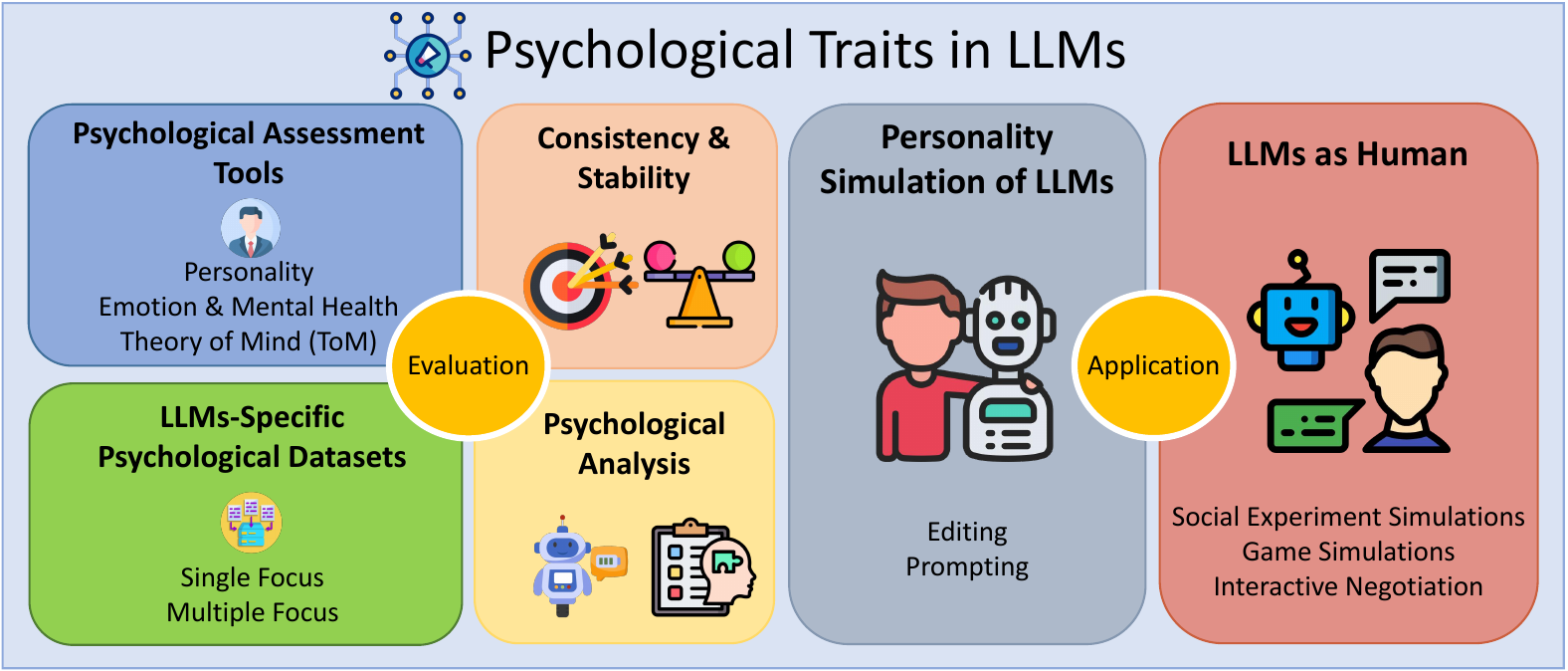} 
  \caption{Overview of Psychological Traits and Human Simulations in LLMs.}
  \label{LLM_Psych_Overview_Figure}
\end{figure*}
\begin{abstract}
As large language models (LLMs) are increasingly used in human-centered tasks, assessing their psychological traits is crucial for understanding their social impact and ensuring trustworthy AI alignment.
While existing reviews have covered some aspects of related research, several important areas have not been systematically discussed, including detailed discussions of diverse psychological tests, LLM-specific psychological datasets, and the applications of LLMs with psychological traits.
To fill this gap, we systematically review six key dimensions of applying psychological theories to LLMs: (1) assessment tools; (2) LLM-specific datasets; (3) evaluation metrics (consistency and stability); (4) empirical findings; (5) personality simulation methods; and (6) LLM-based behavior simulation.
Our analysis highlights both the strengths and limitations of current methods.
While some LLMs exhibit reproducible personality patterns under specific prompting schemes, significant variability remains across tasks and settings.
Recognizing methodological challenges such as mismatches between psychological tools and LLMs' capabilities, as well as inconsistencies in evaluation practices, this study aims to propose future directions for developing more interpretable, robust, and generalizable psychological assessment frameworks for LLMs.
\end{abstract}

\section{Introduction}
In recent years, large language models (LLMs) have made significant progress in natural language processing and artificial intelligence.
Notable examples include OpenAI's GPT-3~\cite{brown2020language} and GPT-4~\cite{ouyang2022training}, as well as Meta's LLaMA series~\cite{touvron2023LLaMA}, Anthropic's Claude~\cite{claude2025}, and Google's PaLM~\cite{chowdhery2022palm}.
These models are trained using large-scale data and have demonstrated capabilities surpassing traditional models in language generation~\cite{zhao2023survey} and reasoning~\cite{DBLP:journals/fcsc/ZhangWDZC25}.

With the growing adoption of LLMs, particularly in psychology and social interactions, they show impressive potential in diverse applications such as mental health support~\cite{stade2024large}, educational tutoring~\cite{bakas2023integrating}, and social dialogue~\cite{ou2023dialogbench}.
Their psychological traits may subtly influence users' perceptions and behaviors, particularly among adolescents.
Therefore, it is crucial to comprehensively evaluate these models' psychological characteristics to understand their potential social impact on human-like interactions.

Psychological assessment tools, such as the Myers-Briggs Type Indicator (MBTI)~\cite{briggs1976myers}, Big Five Inventory (BFI)~\cite{john1991big,john1999big}, and the Short Dark Triad (SD-3)~\cite{furnham2013dark}, have been widely used to evaluate human personality traits and social behaviors.
These tools have been adapted to evaluate LLMs' behavioral patterns and their similarity to human traits across various tasks~\cite{singh2023can,huang2023chatgpt,song2024identifying,DBLP:journals/corr/abs-2406-14703,li2024quantifying,zhang2024better,bodrovza2024personality,la2024open}.
By applying these tools, researchers can explore LLMs' personality dimensions, including their consistent behavioral tendencies and specific personality dimensions~\cite{miotto2022gpt,karra2022estimating,song2023have,singh2023can,DBLP:conf/nips/JiangXZHZ023}.

Existing research (as shown in~\Cref{LLM_Traits}) has shown that LLMs exhibit capabilities similar to human performance in certain psychological tasks.
For example, some LLMs exhibit performance in Theory of Mind (ToM) tasks comparable to that of young children~\cite{van2023theory}.
ToM refers to the ability to understand others' beliefs, intentions, and emotions, which is crucial for human social behavior~\cite{premack1978does}.
However, in complex high-order social reasoning tasks, LLMs still exhibit significant shortcomings~\cite{kosinski2023theory,he2023hi}.
For instance, LLMs struggle with non-literal language, such as sarcasm and metaphor, or scenarios involving complex mental state attributions. 
Their performance significantly lags behind adults~\cite{he2023hi}.

In this work, we present a comprehensive review that systematically examines the psychological characteristics of LLMs.
By synthesizing existing research, we aim to summarize LLMs' psychological characteristics, their performance in human-like interactions, and the associated challenges in psychological assessment.
The structure of the review is illustrated in~\Cref{LLM_Psych_Overview_Figure}.

Our key contributions are summarized as follows:
\begin{enumerate}
    \item \textbf{Systematic Analysis of Psychological Assessment Tools Applied to LLMs}: We provide a comprehensive analysis of how psychological assessment tools have been adapted and utilized to evaluate LLMs. 
    We objectively examine the employed methodologies and assess their effectiveness in measuring psychological attributes of LLMs, such as personality traits, emotional intelligence, and ToM.
    
    \item \textbf{Comprehensive Review of Specialized Datasets and Model Capabilities}: We survey specialized datasets designed to assess various psychological characteristics of LLMs, including personality, emotional abilities, and social cognition. 
    We further analyze the performance of LLMs on these datasets, offering insights into their strengths and limitations in simulating human cognitive and emotional processes.
    
    \item \textbf{Exploration of Human Role Simulations and Identification of Research Gaps}: We explore the capacity of LLMs to simulate human roles across diverse scenarios, categorizing anthropomorphic behaviors into three types: social experiment simulations, game simulations, and interactive negotiation. 
    We also identify critical methodological gaps in the psychological evaluation of LLMs and propose recommendations to improve the validity and reliability of future research.
\end{enumerate}

\section{Psychological Assessment Tools}
\begin{figure}[htbp]
  \centering
  \includegraphics[width=\linewidth]{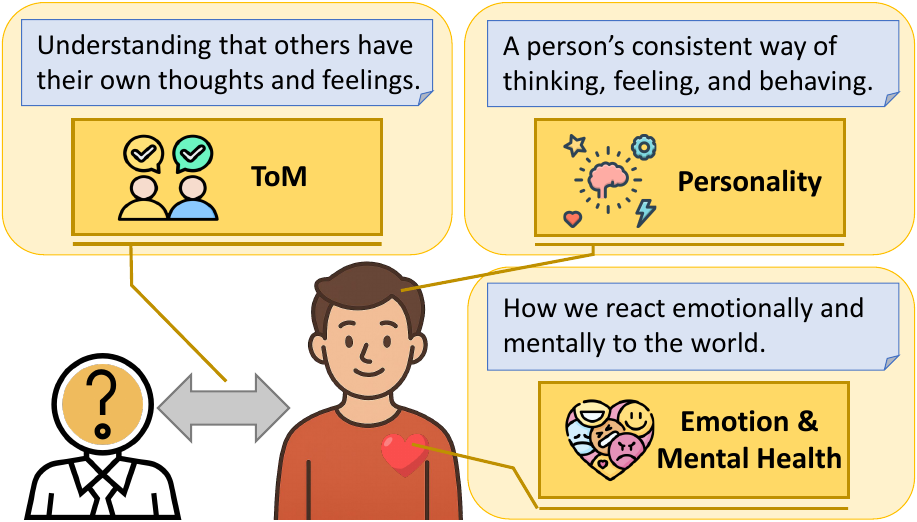} 
  \caption{An illustration showing the relationship between Personality, Emotion \& Mental Health, and Theory of Mind.}
  \label{illustration}
\end{figure}

This section briefly discusses the traditional psychological assessment tools employed in previous studies. 
As shown in~\Cref{illustration}, we categorize these assessment tools into three types: personality, emotion, mental health, and ToM. 
The upper section of~\Cref{classification} provides an overview of the key points discussed in this chapter.

\subsection{Personality Measurement}

Personality measurement uses tools and methods to assess individual traits and behaviors.

\textbf{MBTI}
assesses personality types based on Carl Jung's psychological type theory. 
This instrument categorizes individuals into one of 16 personality types using four dimensions: Extraversion-Introversion, Sensing-Intuition, Thinking-Feeling, and Judging-Perceiving~\cite{briggs1976myers}. 
The primary aim of the MBTI is to help individuals understand their preferences, thereby enabling them to make more informed decisions about their careers and lives. 
Its applications span across business, education, and personal development domains~\cite{articleMBTI}.

At its core, the MBTI seeks to implement Jung's theory of types, which posits that many seemingly random variations in human behavior are orderly and consistent, stemming from fundamental differences in how people perceive and judge. 
This theoretical foundation provides a unique perspective on personality, particularly suited for understanding decision-making processes and interaction styles~\cite{myers1962myers}.

Methodologically, the MBTI employs a forced choice (yes/no) item format, requiring respondents to choose between two options for each question~\cite{briggs1976myers}. 
This approach is designed to reveal an individual's dominant preferences rather than scoring on a continuous scale. 
Through this forced choice mechanism, the MBTI attempts to capture fundamental differences in individuals' perception and judgment preferences~\cite{myers1985guide}.

The MBTI scale applies to most LLMs due to its relatively low reading comprehension requirement, which is equivalent to a seventh-grade level. 
LLMs that have undergone extensive text-based training are typically capable of completing tasks at this level of complexity~\cite{zhang2024better}.

\textbf{BFI}
is a self-report scale that is designed to measure the big five personality traits. 
It consists of 44 brief statements designed to evaluate five fundamental personality dimensions: Openness, Conscientiousness, Extraversion, Agreeableness, and Neuroticism (collectively known as the OCEAN model)~\cite{john1999big}. 
Instead of single adjectives, BFI items employ one or two prototypical trait adjectives as their core, supplemented with elaborative or contextual information to enhance response consistency.

Participants rate each item on a 5-point Likert scale, ranging from 1 (``strongly disagree'') to 5 (``strongly agree'')~\cite{john1999big}. 
Scores for each dimension are calculated by averaging the ratings of all items within that dimension. 
Despite its conciseness, the BFI maintains good content coverage and psychometric properties, making it an effective instrument for efficient and flexible personality assessment when more nuanced trait measurement is not required~\cite{john1999big}.

The BFI is widely used in various research domains, including psychology, education, and organizational behavior. 
In recent years, it has also been utilized to assess the personality characteristics of LLMs~\cite{serapio2023personality, li2022does, singh2023can, DBLP:conf/nips/JiangXZHZ023, jiang2023personallm, huang2023revisiting}.

\textbf{International Personality Item Pool - Neuroticism, Extraversion and Openness (IPIP-NEO)}
is a widely used open-source self-reported questionnaire that assesses personality based on the OCEAN Model. 
The original IPIP-NEO contains 300 questions~\cite{goldberg1999broad}. 
To simplify the test process, \citet{johnson2014measuring} develop a 120-item version IPIP-NEO and demonstrate that its psychometric properties are comparable to the 300-item long version. 
Similarly, \citet{Kajonius2019AssessingTS} and \citet{J2019Using} develop their own 120-item version and 60-item version instruments, respectively.

The validity of International Personality Item Pool(IPIP) is established in different populations, including Greek~\cite{M2015Psychometric} and Romanian samples~\cite{S2012Evaluarea}, and diverse cultural contexts, including Malaysia~\cite{N2021Initial} and Nigeria~\cite{undefined2012Reliability}. 
Furthermore, \citet{Beng2006Assessing} find support for the IPIP's construct validity when compared to the NEO-FFI. 
These studies affirm the IPIP's utility as a robust, accessible personality assessment tool across various cultural settings and item lengths.

\textbf{SD-3}
is composed of three distinct traits: Machiavellianism (a manipulative attitude), Narcissism (excessive self-love), and Psychopathy (lack of empathy). 
While distinct, these three traits collectively represent the darker aspects of human nature, characterized by a core of empathy deficits and manipulative tendencies. 
Studies have linked these traits to a range of adverse behaviors, including bullying, fraud, and criminal activities~\cite{furnham2013dark}.

To assess these traits more efficiently, researchers developed the SD-3 scale~\cite{jones2014introducing}. 
This compact instrument, consisting of 27 items, aims to measure the three traits comprehensively. 
Validated across diverse populations, the SD-3 has exhibited strong reliability and validity, with findings that closely correspond to those obtained using traditional, more extensive measurement tools.

\textbf{HEXACO}
is a six-dimensional personality model: Honesty-Humility, Emotionality, Extraversion, Agreeableness, Conscientiousness, and Openness to Experience~\cite{ashton2009hexaco}. 
This comprehensive framework gauges individual personality characteristics through the use of self-administered questionnaires. 
Various versions of the HEXACO questionnaire have been developed, with the 60-item and 100-item versions being particularly noteworthy. 
Research demonstrates that the 60-item version exhibits psychometric qualities, rendering it a favored choice among researchers in numerous studies~\cite{ashton2009hexaco,agreement2019meta}. 
During the assessment, participants are asked to indicate their level of agreement with a variety of statements. 
Responses are measured on a five-point scale, where 1 represents strong disagreement and 5 denotes strong agreement. 
This process yields composite scores for each of the six dimensions, each of which has 10 distinct items.

\textbf{The Short Scale for Creative Self (SSCS)}
is a response to a call for more elaborate measures of creative self-efficacy (CSE). 
The included statements are based on the concept of general creative self-efficacy within the creative self-concept. 
It consists of 11 items, of which six are considered to measure CSE, four measure creative personal identity (CPI), and one assesses self-rated creativity~\cite{karwowski2018measuring}.

Recent works show that SSCS has multiple characteristics:
First, SSCS correlates with other creativity measures, for instance, the Barron Welsh Art Scale~\cite{furnham2005relationship}. 
Second, although SSCS assesses personality traits and emotional capabilities, since these self-assessments influence creative self-efficacy, it also reflects motivation and willingness to engage in creative behaviors~\cite{kaufman2010american, karwowski2011doesn}. 
Third, SSCS shows domain specificity, meaning that individuals may assess their creativity differently across various domains such as art, science, or everyday problem-solving~\cite{kaufman2010american}.

\subsection{Emotional \& Mental Health Measurement}

Emotional \& mental health measurement aims to assess Emotional well-being and psychological states.

\textbf{Positive and Negative Affect Schedule (PANAS)}
is a widely utilized instrument for assessing individual differences in emotions.
It measures the Positive Affect (PA), which reflects feelings of enthusiasm, alertness, and activity, and Negative Affect (NA), which represents subjective distress and unpleasant engagement~\cite{watson1988development}, respectively.
The PANAS has demonstrated strong reliability and validity across various populations, time frames, and cultural settings~\cite{Lim2010Positive, Petraskaite2022Development, Serafini2016Psychometric}.
It exhibits high internal consistency and good test-retest reliability.
The two dimensions are largely independent, allowing for separate measurement of positive and negative emotions.
Evidence consistently supports that PA and NA are distinct but related constructs~\cite{Rush2014Differences, Carvalho2013Structural}.

In the meantime, some researchers propose alternative factor structures to measure the emotions.
For instance, a three-factor model is suggested to better capture the complexity of affective experiences within individuals, especially in daily life or experience-sampling contexts~\cite{Eadeh2020Multigroup, Cooke2022Examining}.
While the PANAS has proven robust across diverse cultural contexts, \citet{Lee2019Examining} highlight specific items that may be interpreted differently depending on linguistic or cultural norms, which can affect the comparability of results across diverse groups and raise concerns about cross-cultural measurement invariance.

The PANAS is effectively used to study mood fluctuations related to daily activities, stress, and circadian rhythms, as well as in clinical settings, such as monitoring emotional states in individuals undergoing treatment for substance use disorders~\cite{Serafini2016Psychometric}.
Despite its versatility, researchers and practitioners are advised to exercise caution when applying the PANAS across culturally diverse populations.
Attention should be given to possible variations in factor structures, item meanings, and response patterns that may arise due to cultural or contextual influences.
Its brevity and robust psychometric properties make it a valuable instrument for researchers across various disciplines interested in measuring emotional states, but tailoring the interpretation and, if necessary, adapting the instrument may enhance its effectiveness and accuracy in cross-cultural research and practice.

\textbf{Buss-Perry Aggression Questionnaire (BPAQ)}
is a 29-item self-report measure designed to assess aggressive behaviors across physical aggression, verbal aggression, anger, and hostility~\cite{buss1992aggression}.
Although the BPAQ has been extensively validated, most studies have focused on narrow populations such as college students.
To address this limitation,~\citet{gerevich2007generalizability} examine its psychometric properties in a nationally representative Hungarian adult sample to evaluate its generalizability.
Their findings largely support the original four-factor structure, with Physical Aggression, Hostility, and Verbal Aggression factors showing good replication, while the Anger factor replicated moderately well.
This research finds higher scores for males on Physical and Verbal Aggression,consistent with the previous research.
Furthermore, the BPAQ has been validated and applied across various demographic groups and cultural contexts, confirming its structural integrity and psychometric soundness~\cite{Cunha2021Buss, Reyna2011Buss, Castrillon2009Cualidades}.
For example,~\citet{Cunha2021Buss} demonstrate its applicability in Portuguese men convicted of intimate partner violence, while~\citet{Reyna2011Buss} confirm its construct validity among Argentinean adolescents.
Similarly,~\citet{Castrillon2009Cualidades} verify the instrument’s utility among Colombian university students, suggesting its sensitivity to culturally embedded expressions of aggression.
In response to the need for a more efficient assessment, a 12-item short form of the BPAQ (BPAQ-SF) was developed, retaining the original four-factor structure~\cite{Webster2013brief, Christopher2024Psychometric}.
The short form demonstrates strong psychometric properties, including high internal consistency and robust construct validity, making it particularly suitable for large-scale surveys and clinical contexts where brevity is essential.

The BPAQ is translated and validated across multiple languages, attesting to its cross-cultural applicability.
For instance,~\citet{Tamyres2020Questionario} validate the Portuguese short version through structural analyses, while~\citet{Abd2020CONTENT} establish the content validity and reliability of the Malay adaptation, employing rigorous methodologies such as forward-backward translation and confirmatory factor analysis to ensure cross-cultural robustness.
Research has shown significant associations between BPAQ scores and variables such as gender, education, and psychiatric symptoms~\cite{Jeyagurunathan2022Aggression}.
\citet{Jeyagurunathan2022Aggression} find that higher symptom severity in individuals with schizophrenia correlated with increased aggression scores on BPAQ.
These findings highlight the instrument’s relevance across general, clinical, and forensic populations.

Both BPAQ and BPAQ-SF demonstrate strong psychometric qualities across diverse populations and cultural settings.
Their structural validity, reliability, and adaptability to various languages make them robust tools for assessing aggression in research and clinical practice.
Nevertheless, researchers must remain mindful of cultural and contextual factors that may influence respondents’ interpretation and response behavior.

\begin{figure*}[htbp]
  \centering
  \includegraphics[width=0.8\textwidth]{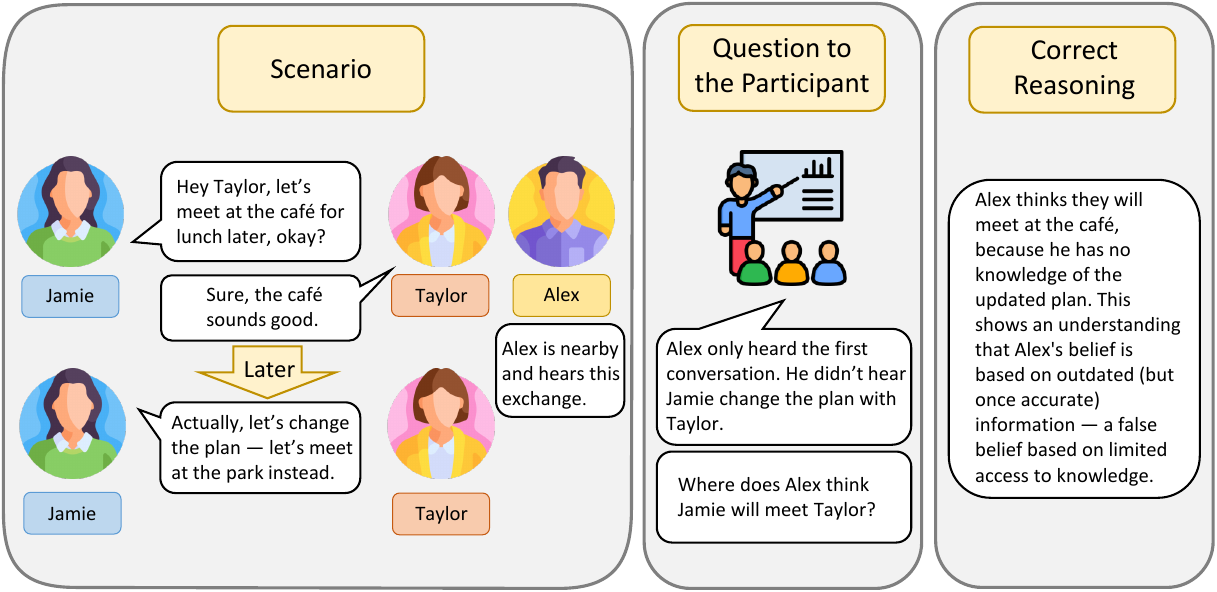} 
  \caption{An example of Imposing Memory Test.}
  \label{Imposing_memeory}
\end{figure*}

\subsection{ToM}

ToM refers to the ability to comprehend one's own mental states, as well as those of others, facilitating the prediction of their actions in specific situations based on such understanding~\cite{schlinger2009theory}.

\textbf{False Belief Tests}, also known as the unexpected transfer task, are classic psychological assessments used to evaluate an individual's ToM. 
The primary objective of these tasks is to determine whether a participant can understand that others may hold false beliefs and can distinguish between those beliefs and actual reality~\cite{wimmer1983beliefs}. 
These tests consist of paradigmatic tasks such as the ``Sally-Anne Test'' and the ``Smarties Test'', both of which are pivotal in probing an individual's capacity to understand that others may hold beliefs that diverge from reality due to incomplete or erroneous information.

\begin{enumerate}
    \item \textbf{Sally-Anne Test}~\cite{baron1985does}: This task evaluates an individual's ability to recognize that a character, Sally, may hold a false belief about the location of an object (e.g., a ball), based on her limited access to situational updates. 
    The test is structured to measure both first-order beliefs and second-order beliefs. 
    \textbf{First-order beliefs} involve understanding what another person believes about the world, such as Sally believing that the ball is still in the basket where she placed it. 
    \textbf{Second-order beliefs} involve understanding one character's belief about another character's belief. 
    For example, in an extended version of the Sally-Anne test, Sally and Anne are in a park where an ice cream vendor initially stands by a fountain. 
    Anne goes to get her wallet, and while she is away, the vendor moves to the swings but informs Sally of his new location. 
    Later, the vendor also tells Anne of his new position. 
    The question then is, ``Where does Sally think Anne will go to buy ice cream?'' The correct answer is the fountain, as Sally does not know that Anne has also been informed of the new location~\cite{perner1985john}. 
    Such assessments are critical for determining the developmental trajectory of mental state attribution.
    \item \textbf{Smarties Test}~\cite{perner1987three}: The Smarties Test investigates whether participants can understand that others may have incorrect beliefs about the content of a container, based solely on external cues. 
    In this task, participants must infer that a character will be misled by the label on the container, which does not match its actual contents. 
    This test serves as a crucial measure of an individual’s ability to reconcile conflicting perspectives, particularly when they diverge from known reality.
\end{enumerate}

\textbf{Strange Stories Test}
comprises a series of social scenarios that are designed to assess nuanced aspects of ToM, including understanding of non-literal language, such as lying, sarcasm, and irony. 
Participants are required to explain the mental states, motivations, and intentions of characters depicted in these scenarios~\cite{happe1994advanced, kaland2005strange}. 
For example, consider a scenario where a child receives a box that she believes contains her favorite toy, but it turns out to be books. 
The child then smiles and says she loves the gift. 
The participants are asked if the child really means what she says and why she might have said it. 
The correct answer would reveal that the child is pretending to like the gift to avoid hurting the feelings of the person who gave it to her. 
The increasing complexity of the scenarios necessitates sophisticated social reasoning and the ability to interpret indirect communicative cues, thus providing insights into higher-level ToM abilities.

\textbf{Imposing Memory Test}
is modified for applicability to children aged 7-10, evaluating both the inferential and mnemonic aspects of Theory of Mind~\cite{van2023theory, kinderman1998theory, duijn2016lazy}. 
This test is particularly focused on assessing recursive reasoning abilities regarding mental states and the participant's capacity for factual recall within social contexts. 
For example, as shown in ~\Cref{Imposing_memeory}, participants are asked to interpret scenarios involving multiple layers of mental state attribution. 
Consider a scenario where a child, Alex, observes two friends, Jamie and Taylor, discussing where to meet for lunch. 
Jamie initially tells Taylor that they will meet at the café. 
However, Jamie changes the location to the park but only informs Taylor. 
In the absence of access to the second conversation, Alex believes that Jamie still thinks they will meet at the café. 
The participants must determine what Alex believes about Jamie and Taylor's plans. 
This type of task assesses the participants' understanding of nested beliefs and helps reveal their ability to engage in higher-level recursive thinking~\cite{van2023theory}. 
Such tasks provide insights into the cognitive mechanisms underpinning complex belief attribution and their developmental progression.

\textbf{Faux Pas Test}
is designed to assess individuals' ability to recognize inappropriate remarks in social situations and the underlying mental states. 
Participants are presented with a series of short stories depicting social interactions. 
In each story, a character (the speaker) unintentionally makes an offensive remark without realizing its inappropriateness~\cite{baron1999recognition}. 
Following the story presentation, questions are posed to test the participants' understanding of the speaker's false beliefs, exploring their grasp of the speaker's mental state. 
Understanding a faux pas scenario requires comprehending two mental states: the speaker's unawareness of the inappropriateness of their words and the listener's (the victim's) potential emotional reaction to those words. 
This task enables researchers to assess participants' ability to integrate information within stories, accurately infer the mental state of the speaker, and thus assess their understanding of complex social situations.

\subsection{Summary}
Previous studies relies on various psychological assessment tools to evaluate LLMs, focusing on their cognitive, emotional, and social capabilities similar to human assessment. 

Specifically, personality measurement tools assess diverse personality dimensions, ranging from basic traits like extraversion and conscientiousness to complex constructs such as creative self-concept and dark personality traits. 
Notably, their item formats vary from forced-choice to Likert scales, offering both breadth and specificity in assessment. 
Consequently, many of these tools, such as the BFI and MBTI, have been used to evaluate the personality profiles of LLMs due to their manageable reading comprehension levels and strong psychometric properties.

Similarly, emotional and mental health measurement tools utilize self-report questionnaires to capture individuals' internal affective states. 
For instance, PANAS and BPAQ measure emotional well-being, aggression, and mood regulation. 
These instruments are valued for their ability to capture dynamic emotional patterns and provide insights into how emotional responses may fluctuate across different contexts. 
Furthermore, their reliability and cultural adaptability make them suitable for evaluating LLMs’ responses in emotionally charged or socially sensitive scenarios.

In addition, ToM assessment focuses on an individual's capacity to attribute and reason about mental states, an ability considered foundational to social cognition. 
In this domain, tools such as the False Belief Tests, Strange Stories Test, Faux Pas Test, and Imposing Memory Test probe different levels of ToM reasoning, from basic Belief attribution to complex recursive thinking and recognition of social faux pas. 
As a result, these tasks are increasingly employed in LLM evaluations to determine whether such models can accurately infer others’ beliefs, intentions, or emotional states from narrative or conversational contexts.

Therefore, these tools offer a multi-dimensional framework for evaluating LLMs, including personality structure, emotional regulation, and social reasoning. 
Their adaptation reflects a broader trend toward modeling artificial intelligence in line with human psychological constructs, enhancing our understanding of machine capabilities and the boundaries between artificial and human cognition.

\begin{figure*}[htbp]
  \centering
  \includegraphics[width=0.8\textwidth]{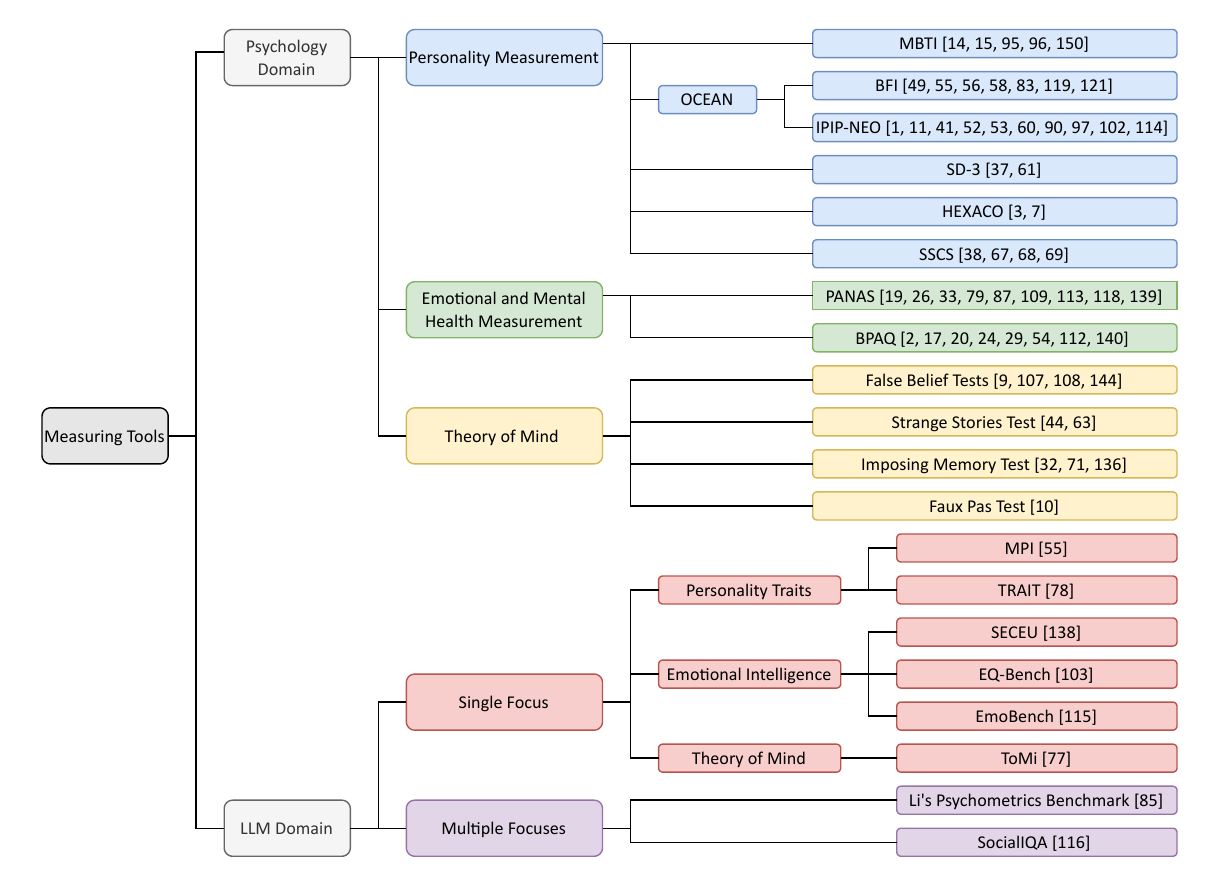} 
  \caption{Datasets Classified by Category and Domain.}
  \label{classification}
\end{figure*}

\section{LLMs-Specific Psychological Datasets}

In this section, we focus on examining the datasets developed in recent years for evaluating LLMs.
These innovative datasets are informed by established psychological frameworks while integrating model-specific considerations, thereby facilitating a more rigorous and nuanced evaluation of LLMs across multiple psychological dimensions.
The lower part of~\Cref{classification} presents a summary of this chapter.

\subsection{Personality}

Datasets designed to measure the personality of LLMs typically consist of multiple personality traits assessment tools, employing different theories to evaluate the personality of LLMs from various perspectives.

\textbf{Machine Personality Inventory}~(MPI)~\cite{DBLP:conf/nips/JiangXZHZ023} is a suite of multiple choice questions based on the theory of the Big Five personality traits, designed to evaluate the behavior of LLMs from a personality perspective quantitatively. 
The construction of MPI items draws from the International Personality Item Pool (IPIP) and its derived versions~\cite{goldberg1999broad, goldberg2006international, johnson2005ascertaining, johnson2014measuring}, as well as the short 15-item Big Five Inventory~\cite{lang2011short}. 
Each test item within the MPI consists of a question paired with a set of response options, aiming to assess the model's suitability regarding a particular self-descriptive statement, with the model being required to select one of the given answers. 
All elements are annotated according to the five dimensions of the Big Five personality traits, ensuring coverage across different personality characteristics.

The MPI uses a Likert scale (ranging from ``Very Accurate'' to ``Very Inaccurate'') to score each model's response. 
By methodically aggregating scores across all test items, the model's overall scores in the Big Five personality dimensions, referred to as the OCEAN scores, can be calculated. 
These quantitative scores range from 1 to 5, reflecting the model's inclination towards each personality dimension. 
Moreover, MPI evaluates the stability of the model's ``personality'' by examining its internal consistency: how consistently the model responds to different questions targeting the same personality dimension. 
A model providing consistent answers across questions related to the same personality dimension indicates a stable personality characteristic.

\textbf{TRAIT}~\cite{DBLP:journals/corr/abs-2406-14703} is a novel personality assessment tool designed explicitly to evaluate LLMs, which addresses the limitations of conventional self-assessment personality tests in terms of validity and reliability. 
With 8,000 multiple-choice questions, TRAIT assesses LLMs across personality dimensions based on BFI and SD-3. 
By integrating the ATOMIC10× commonsense knowledge graph~\cite{DBLP:conf/naacl/WestBHHJBLWC22}, TRAIT expands the description of personalities into a wide range of real-world contexts, encompassing various physical and social scenarios, providing a more authentic assessment of LLMs behavior patterns. 
Compared to traditional self-assessment approaches, TRAIT offers superior validity and reliability, overcoming inconsistencies arising from the subjective nature of self-reporting~\cite{DBLP:journals/corr/abs-2406-14703}.

The construction of TRAIT utilizes a collaborative human-AI approach. Initially, GPT-4 expands a set of self-assessment questionnaires into 1,600 diversified personality descriptions. 
Subsequently, the ATOMIC10× knowledge graph is employed to extract the most relevant scenarios associated with these descriptions, resulting in 8,000 detailed situations.
Each scenario is paired with four multiple-choice options, allowing for comprehensive capture of various aspects of personality traits. 
Moreover, TRAIT undergoes human evaluation by psychological professionals, who review a random sample of 200 items, achieving an accuracy rate of 97.5\%, thus validating the quality of the dataset. 
Eventually, TRAIT significantly reduces refusal rates in personality assessment, while demonstrating high robustness in sensitivity to prompts and option order, ensuring a reliable and consistent measurement process.

\subsection{Emotional Abilities}
Emotional Understanding (EU), a core component of Emotional Intelligence (EI), refers to an individual's ability to recognize, interpret, and comprehend emotions within social contexts. 
To assess this capability, researchers develop various testing tools, including Situational Evaluation of Complex Emotional Understanding (SECEU)~\cite{wang2023emotional}, EQ-Bench~\cite{DBLP:journals/corr/abs-2312-06281}, and EmoBench~\cite{DBLP:conf/acl/Sabour0ZLZSLMH24}.

\textbf{SECEU} is a standardized psychometric instrument specifically designed to assess EU capacity. 
It comprises 40 situational items, each delineating a complex scenario set in academic, familial, or social environments, engineered to elicit a blend of positive and negative emotional responses. 
Participants assess the intensity of four emotions (e.g., surprise, joy, puzzlement, pride) for each scenario, allocating a total of 10 points across these emotions. 
SECEU uses a consensus scoring method for standardization, with normative data derived from a sample of 541 university students (mean age = 22.33 years, SD = 2.49). The instrument shows robust internal consistency (Cronbach's $\alpha$ = 0.94) and validity. 
Raw scores are converted into standardized Emotional Quotient (EQ) scores, with a mean of 100 and a standard deviation of 15.

\textbf{EQ-Bench} is a newly developed benchmark with 60 English questions, specifically designed to evaluate the emotional intelligence of LLMs.
Each question features a dialogue scenario depicting conflict or tension, along with four emotions to be rated. 
Similar to SECEU, EQ-Bench requires models to assess the emotional intensity of specific characters, but expands the rating scale to 0-10. 
EQ-Bench incorporates several improvements over SECEU: it uses more complex dialogue scenarios, selects a more diverse range of emotions for rating, employs expert-chosen reference answers rather than crowd averages, and removes the constraint on the sum of emotional intensities.

\textbf{EmoBench} extends the assessment of Emotional Intelligence in LLMs abd consists of 400 hand-crafted multiple-choice questions available in English and Chinese.
It introduces a comprehensive framework for evaluating both EU and Emotional Application (EA), aiming to transcend simple pattern recognition, necessitating reasoning and understanding of emotional implications~\cite{DBLP:conf/acl/Sabour0ZLZSLMH24}.
The EU component assesses the model's ability to identify emotions and their causes across four categories: complex emotions, personal beliefs and experiences, emotional cues, and perspective-taking. 
The EA component evaluates the model's proficiency in leveraging emotional understanding to identify the most effective response or action within emotional dilemmas involving personal and social relationships.

EmoBench supplies a challenging evaluation of emotional intelligence, as evidenced by the performance gap between current language models and human participants. 
The best-performing model (GPT-4) achieved an accuracy of 59.75\% and 75.88\% on the EU and EA tasks, however, which is even lower than the average performance of humans. 
EmoBench's results suggest that existing LLMs still struggle with emotional intelligence, particularly in understanding complex emotional scenarios.

\subsection{ToM}
The ToMi dataset~\cite{le2019revisiting} is designed to evaluate the ability of AI systems to understand ToM. 
The construction is based on classic psychological tests, such as the Sally-Anne test and other experiments used to evaluate higher-order beliefs. 
By automatically generating stories and related question-answer pairs, the ToMi dataset simulates the mental states of different agents, allowing models to infer the intentions and beliefs of agents during natural language dialogue. 
Compared to traditional evaluation methods, the ToMi dataset places particular emphasis on controlling systematic biases in the data generation process. 
By adding random distractors, it enhances the assessment of the model's generalization capabilities, preventing models from making inferences solely based on inherent regularities in the data rather than true ToM reasoning.

\subsection{Comprehensive}
The psychometrics benchmark introduced by~\citet{li2024quantifying} provide a rigorous and nuanced framework for assessing the psychological attributes and behaviors of LLMs. 
This benchmark addresses six core psychological dimensions: personality, values, emotion, ToM, motivation, and intelligence. 
Utilizing thirteen diverse datasets that span a wide range of scenarios and item types, the benchmark facilitates a systematic and quantitative analysis of LLMs within established psychological frameworks. 
In contrast to traditional evaluations that predominantly focus on assessing capabilities, this benchmark employs a variety of assessment methods, including self-reports, open-ended questions, and multiple-choice formats, to expose latent discrepancies between LLMs' self-perceived traits and their actual exhibited behaviors. 
These inconsistencies mirror social desirability biases commonly observed in human respondents, suggesting that LLMs may also generate responses that deviate from their actual behavior in more open-ended, less structured contexts.

The framework, as proposed by~\citet{li2024quantifying}, adheres to a psychometric methodology encompassing the identification of psychological dimensions, dataset curation, and meticulous evaluation with comprehensive validation of results. 
Reliability is a pivotal component of this framework, with a variety of validation techniques employed to ensure the robustness and interpretability of the outcomes. These comprise internal consistency, parallel forms reliability, inter-rater reliability, option position robustness, and adversarial robustness. 
The evaluations reveal that LLMs exhibit consistent reasoning and behavioral patterns in some contexts, such as emotional comprehension, motivational expression, and ToM tasks. 
However, significant variability and incoherence emerge when LLMs are confronted with ambiguous or emotionally complex scenarios. 
This benchmark thus introduces a novel approach to the psychological evaluation of LLMs, elucidating the nuanced behaviors exhibited across different psychological dimensions and yielding critical insights into both their strengths and their inherent limitations.

SOCIAL IQA~\cite{sap2019socialiqa} is a large-scale multiple-choice question-answering benchmark designed to evaluate the ability of computational models to reason within social contexts. 
The dataset contains 37,588 question-answer pairs, each consisting of a specific social scenario description and three candidate answers, intended to test various aspects of emotional and social intelligence, such as motivation reasoning, emotional reaction inference, and prediction of subsequent actions. 
The SOCIAL IQA dataset was constructed using a crowdsourcing approach, with a multi-stage annotation process that reduces stylistic biases introduced by incorrect answers. 
The dataset is specifically designed to reflect a model's ability to reason about complex social scenarios. 
Human performance on this dataset is approximately 87\%.

\subsection{Summary}
Existing research has created psychological datasets more suitable for measuring LLMs by combining or modifying psychological assessment tools. 
Compared to traditional measurement tools, these datasets aim to provide a systematic analysis of the psychological attributes and behaviors of LLMs, emphasizing the consistency of models. 
They help people gain a deeper understanding of the performance and application of psychology in LLMs.

\section{Consistency and Stability}
\label{sec:consistency_stability}
\begin{figure}[htbp]
  \centering
  \includegraphics[width=0.9\linewidth]{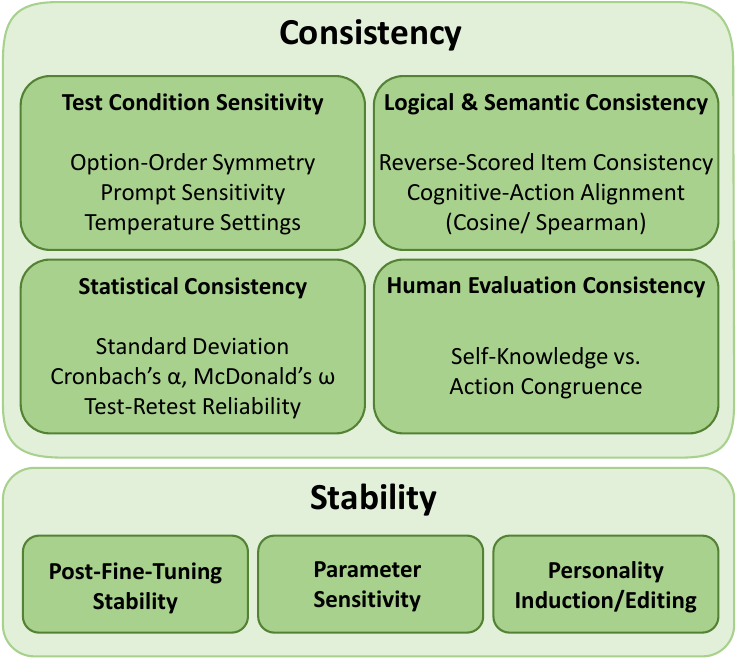} 
  \caption{Key Dimensions in LLMs'  Psychological Assessment: Consistency vs. Stability}
  \label{Consistency_Stability}
\end{figure}

LLMs have become pivotal in psychological research, particularly in evaluating personality traits, cognitive behaviors, and therapeutic interactions.

This section explores the \textit{consistency} and \textit{stability} of LLMs from a psychological standpoint, crucial for their reliability in such applications.
Consistency refers to the models' ability to produce similar outputs when given similar inputs, encompassing internal reliability (e.g., McDonald's $\Omega$ and Intraclass Correlation Coefficient, ICC) and sensitivity to prompt variations, such as option order in psychological tests.

Stability, on the other hand, is less clearly defined in the current studies, but refers to the consistency of their psychological attributes when subjected to fine-tuning or parameter changes.
It examines the persistence of psychological test performance if the parameter of it has been changed (e.g., before and after temperature change or fine-tuning), a process where models are further trained on specific datasets to enhance task-specific capabilities.
The key dimensions for assessing these properties are visualized in~\Cref{Consistency_Stability}.

\subsection{Consistency}
Consistency evaluation is a critical method for assessing the internal coherence of LLMs when simulating human behavior, cognition, and personality traits.
It comprehensively evaluates whether the model's responses are consistent under various conditions through multiple dimensions and technical approaches. 
Below is a comprehensive introduction and classification of consistency evaluation:

\mypara{Test Condition Sensitivity Evaluation}
By modifying test conditions and observing the corresponding behavioral changes in the model, this evaluation assesses its sensitivity to test conditions and stability in dynamic environments.

\citet{song2023have} evaluate LLMs using Option-Order Symmetry, where a model's responses to self-assessment questions should remain consistent regardless of the order of the answer choices.
The study reveals that many LLMs fail to maintain this symmetry, indicating inconsistencies in their responses.
Furthermore, even when a model formally preserves option-order symmetry, its answers remain unchanged across different contexts, suggesting insufficient sensitivity to prompt content. 
\citet{gupta2023self} also explore the influence of option order and finds that, in most cases, nearly all models exhibit statistically significant differences in scores based on the order of the options. 
This highlights the substantial impact of option order on model responses, underscoring the challenges of reliably assessing LLMs' personality traits using current self-assessment methods. 
Similarly, \citet{DBLP:journals/corr/abs-2406-14703} examine option-order sensitivity by altering the sequence of options to determine whether this affects LLMs' responses. 
The findings show that TRAIT, a personality assessment tool designed for LLMs, exhibits low sensitivity to option order, with LLM responses remaining consistent across different sequences of options.

\citet{sorokovikova2024llms} investigate the impact of different prompting methods on model evaluation, comparing a standard prompt with a modified version that includes the phrase ``Answer as if you were a person.'' 
In the Big Five personality test, these two prompt variations yielded different model responses. 
For the LLaMA2 model, certain questions could not be answered due to restriction mechanisms when the phrase was absent, whereas including the phrase allowed the model to respond smoothly. 
This suggests that subtle changes in prompts can influence model behavior and, in turn, affect the results of personality assessments, emphasizing the critical role of prompting in shaping how models simulate personality traits. 
\citet{gupta2023self} further investigate prompt sensitivity by comparing model responses to three semantically equivalent prompts from~\cite{DBLP:conf/nips/JiangXZHZ023},~\cite{miotto2022gpt}, and~\cite{huang2023revisiting}, each presenting a Likert scale differently. 
The results indicate that for almost all models and traits, these prompt variations led to significant differences in personality scores, suggesting that model personality assessments are highly sensitive to the phrasing of prompts. 
This raises concerns about the reliability of personality evaluation results in previous studies. 
\citet{DBLP:journals/corr/abs-2406-14703} also evaluate prompt sensitivity by using three different prompt templates from prior research and observing whether LLMs provide consistent responses across three test iterations. 
The findings suggest that TRAIT is less affected by variations in prompt wording.

\citet{sorokovikova2024llms} also evaluate the performance of large language models under different temperature settings, applying various temperature parameters to different models.
Specifically, ChatGPT's temperature settings were 1 (low), 1.5 (medium), and 2 (high), while LLaMA2 and Mixtral had settings of 0.3 (low), 0.7 (medium), and 1 (high). 
The results showed that the Big Five personality scores of the models varied across these temperature conditions. 
In the neuroticism dimension, ChatGPT's scores remained stable with minimal fluctuations, while LLaMA2 showed some variation but remained relatively consistent, and Mixtral exhibited stable scores. 
For other personality dimensions, such as extraversion, openness, agreeableness, and conscientiousness, there were variations in the scores depending on the temperature settings, with some models showing no significant trends. 
These findings indicate that temperature parameters do influence the stability of a model's simulated personality traits, though the degree of impact varies across models. 
\citet{miotto2022gpt} analyze the consistency of GPT-3's responses under different temperature settings, focusing on the stability of personality and value dimension scores. 
In the HEXACO personality assessment, temperature significantly impacted GPT-3’s scores across various dimensions, except for the honesty-humility dimension. Specifically, higher temperatures led to a decrease in emotionality scores, while scores for extraversion, agreeableness, conscientiousness, and openness increased. 
This suggests that GPT-3's personality is not entirely stable across different temperatures, and temperature variations can alter its personality expression. 
In the values assessment, nine out of ten value dimensions showed a significant negative correlation with temperature, indicating that higher temperatures led to lower scores. 
This suggests that, in the absence of response memory, GPT-3's emphasis on most values changes with temperature, resulting in instability in its value assessment scores.

Additionally,~\citet{DBLP:journals/corr/abs-2406-14703} assess Paraphrase Sensitivity, which measures how sensitive LLMs are to changes in the wording of semantically identical questions. 
The study finds that existing self-assessment tests used to evaluate the personality of LLMs perform poorly in terms of paraphrase sensitivity.

\mypara{Statistical Consistency Evaluation}
This approach involves quantifying the stability of a model under different test conditions using statistical metrics to assess the consistency and reliability of its responses.

\citet{li2024evaluating} evaluate the stability of LLMs when responding to the same or similar questions multiple times, measuring consistency by calculating the standard deviation of repeated responses. 
A more minor standard deviation indicates greater stability. 
The study also compares models of different scales to explore the relationship between model size and personality trait consistency. 
The results demonstrate that the same model tends to yield stable scores across identical psychological tests. 
In the SD-3 and BFI tests, the standard deviations of multiple responses from GPT-3, InstructGPT, GPT-3.5, GPT-4, and LLaMA-2-chat-7B were all lower than those observed in different human individuals. 
However, in well-being assessments (FS and SWLS), these models exhibited greater variability, indicating reduced stability compared to the previous tests. 
\citet{zhang2024better} further investigates the stability of repeated evaluations and applies an averaging approach across multiple assessments to mitigate the impact of option order on MBTI evaluation results. 
The study demonstrates that after 30 assessments, all models' MBTI results became consistent across different option orders, confirming that repeated evaluations can enhance measurement stability.

\citet{li2024quantifying} evaluate the internal consistency of LLMs by examining their stability in similar contexts, using the BFI test to measure standard deviations. 
The results indicate that LLaMA3-8B and Mistral-7B exhibit human-like stability with lower standard deviations, while GPT-4 and Mixtral-8×7B show higher variations, particularly in openness, suggesting weaker consistency. 
In the Short Dark Triad test, ChatGPT demonstrates the highest consistency, with standard deviations lower than human averages, whereas other models show greater variability, indicating limitations in stability for dark personality traits. 
For value dimensions, cultural orientation assessments reveal that LLMs are consistent in some dimensions but vary significantly in others. 
For instance, GPT-4's high standard deviation in humanitarian orientation indicates instability due to input variations. 
\citet{huang2023revisiting} further analyze internal consistency by testing LLMs' response stability across 2,500 configurations involving variations in instructions, scale items, language, choice labels, and order. 
The study finds that GPT-3.5-Turbo maintains satisfactory consistency in BFI. 
\citet{DBLP:conf/nips/JiangXZHZ023} assess consistency through MPI scale tests, showing that GPT-3.5 and Alpaca-7B achieve near-human stability across personality traits, further validated by requiring models to explain their answers. 
\citet{petrov2024limited} employ Cronbach’s $\alpha$, Greatest Lower Bound (GLB), and McDonald’s $\omega$ to compare LLMs with human data. 
The study reveals that under general role settings, GPT-3.5 and GPT-4 exhibit acceptable consistency with most $\alpha$ values $ \geq 0.70$.
However, in silicon-based role settings, internal consistency declines sharply, with some $\alpha$ values dropping to 0.10–0.50, indicating reliability issues in personality simulation. 
\citet{serapio2023personality} extend this analysis to IPIP-NEO and BFI subdimensions, showing that instruction-tuned models (e.g., Flan-PaLM 62B) achieve high reliability ($\alpha > 0.90 $), while non-instruction-tuned models (e.g., PaLM 62B) perform poorly ($\alpha$ ranging from -0.55 to 0.67). 
Additionally, larger models within the same training configuration (e.g., Flan-PaLM 8B, 62B, and 540B) demonstrate improved personality assessment reliability, suggesting that both model scale and training methods significantly influence internal consistency.

\citet{bodrovza2024personality} investigate temporal stability, assessing whether LLMs maintain consistent responses to the same psychological measurement tools over time. 
Consistency is evaluated by examining whether LLMs provide stable responses to different questions at the same time point. 
The study compares responses across two time points to evaluate stability and internal consistency reliability in personality assessments such as BFI-2 and HEXACO-100.
Results indicate variations in consistency across models and measurement tools, with LLaMA3 and GPT-4o demonstrating higher consistency, while GPT-4 and Gemini exhibit lower levels. 
Furthermore, consistency is influenced by specific traits, with the Agentic Management dimension showing the highest stability. 
\citet{huang2023revisiting} apply test-retest reliability to measure whether the same assessment yields consistent results for the same group over different periods. 
Using correlation coefficients between two test administrations as reliability indicators, a two-week BFI test on GPT-3.5-Turbo, covering two different model versions, reveals that the mean scores across BFI dimensions remain statistically unchanged after the model update.

In addition, various studies have employed different methodologies to quantify consistency. 
\citet{klinkert2024driving} evaluate LLMs' ability to generate consistent content based on given personality traits. 
Accuracy assessments reveal significant differences between models, with GPT-4-0613 demonstrating superior accuracy and consistency in generating personality-aligned content. 
Root Mean Square Prediction Error (RMSPE) is calculated using Euclidean distance and cosine similarity, with results showing that GPT-4-0613 achieves the lowest RMSPE, surpassing baseline levels and indicating minimal deviation when generating expected personality-driven content. 
Intraclass correlation coefficient (IRR) analysis further confirms GPT-4-0613's superior performance in producing consistent responses, while Linear Discriminant analysis (LDA) reinforces its stability in consistency-based tasks. 
\citet{ai2024cognition} employ split-half reliability, where personality questionnaires are divided into two equal-length parts, and Spearman’s rank correlation coefficient is computed to assess consistency. 
The study finds that ChatGLM3, GPT-3.5-Turbo, GPT-4, Vicuna-13B, and Vicuna-33B perform well in split-half reliability tests, approaching human-level consistency, though LLMs still require improvements in replicating human personality traits reliably. 
\citet{li2024quantifying} examine parallel forms reliability by evaluating LLMs’ response consistency across different test versions. 
Using moral value assessments as a case study, the research alters question formats to analyze model stability. 
Results indicate that in high-ambiguity scenarios, such as the MoralChoice survey, LLMs exhibit reduced consistency across parallel test versions, suggesting that they are more susceptible to prompt variations when facing uncertain or complex questions. 
\citet{frisch2024llm} adopt explicit assessment methods, directly evaluating LLMs' personality traits via BFI testing to determine alignment with predefined personality profiles. 
Findings reveal that creative agents maintain high consistency before and after testing, whereas analytical agents experience trait shifts following interactions, indicating that LLMs' personality expressions may fluctuate due to engagement dynamics.

\mypara{Evaluation of Logical and Semantic Consistency}
By analyzing LLMs’ logical reasoning, semantic comprehension, and behavior generation capabilities, this evaluation assesses their coherence and stability in complex scenarios.

\citet{ai2024cognition} employ logical consistency as an evaluation method to examine the coherence of LLMs when responding to personality questionnaires. 
By designing both positively and negatively scored items, the study assesses whether LLMs carefully read and respond attentively. 
For example, in measuring extraversion, the questionnaire includes both positively phrased statements, such as ``Finish what I start.'' and negatively phrased ones, like ``Leave things unfinished.'' 
The negatively phrased items are reverse-scored to align with the scoring direction. 
If LLMs' responses on a 7-point Likert scale remain statistically consistent between positively and negatively scored items (e.g., both $ \geq 4 $ or $ \leq 4 $), their answers are considered logically consistent. 
The study tested 12 LLMs and identified 7 that provided valid responses, with ChatGLM3, GPT-3.5-turbo, GPT-4, Vicuna-13B, and Vicuna-33B demonstrating outstanding logical consistency. 
This suggests that these models exhibit a level of attentiveness and logical reasoning in personality questionnaires closer to that of humans.
This article also evaluates the performance of LLMs in cognitive-action consistency by designing a bilingual cognitive-action test set. 
The test set includes 180 matching pairs, covering personality cognitive descriptions and real-life action scenarios. 
LLMs are required to assess personality traits in the cognitive questionnaire and choose between two options, A and B, in the scenario questionnaire. 
The answers are mapped to a 1-7 Likert scale. 
By calculating cosine similarity, Spearman rank correlation coefficient, mean difference, and consistency ratio, the article compares the similarity and correlation of LLMs' responses with human responses across the two questionnaires. 
The results show that LLMs exhibit significantly lower cognitive-action consistency than humans, especially in the domain of extraversion, where the consistency ratio is only 17.14\%. 
In contrast, for human participants, cosine similarity and Spearman rank correlation coefficient both exceed 0.75, with a consistency ratio of over 84\%. 
Nevertheless, LLMs perform relatively well in the openness domain, with a consistency ratio of 60\%.

\citet{frisch2024llm} design LLMs' personality traits using prompts.
They use an implicit method to assess whether their language use in generated texts (e.g., personal stories), like personal stories aligns with the designated personality traits.
The study employs Linguistic Inquiry and Word Count (LIWC) software to quantitatively analyze the text generated by the agents.
Their results indicate significant linguistic differences between creative and analytical agents.
Language alignment becomes evident during interactions.

\citet{miotto2022gpt} investigate the impact of response memory on GPT-3's performance in value assessment.
A response memory is used to simulate how human participants would typically remember their responses to previous items.
The experiment modifies prompt structure and adds response memory to compare GPT-3’s consistency, alignment with theoretical models, and deviations from human data, with and without response memory. 
The results show that incorporating response memory enhances response consistency, lowers score variation across value dimensions, and reduces extreme values. 
Furthermore, GPT-3’s responses also align better with the Human Value Survey (HVS) model, as scores within the same value category become more similar, matching the expected value classification.

\citet{liu2024dynamic} evaluate model stability and logical consistency by presenting LLMs with the same question multiple times and analyzing whether their responses contain contradictions. 
If two or more answers differ in meaning, they are considered contradictory. 
The experiment employs multiple response iterations to assess the consistency of the models in both logic and content. 
The results reveal variation in performance across different models. 
For example, in a scripted dialogue test, LLaMA-7B trained with the DPG method outperforms those trained with Freeze-SFT and LoRA-SFT in response consistency. 
This indicates that DPG training enables more stable and logically coherent responses, maintaining higher consistency across different responses.

\mypara{Human Evaluation Consistency} 
This part explores the issues of self-knowledge and action consistency in LLMs, focusing on the relationship between their responses to personality questionnaires and actual behavioral tendencies.
Specifically, it investigates whether the personality traits reflected in their questionnaire responses align with their behavioral tendencies in simulated real-world scenarios.

\citet{ai2024cognition} design a bilingual test set consisting of a personality knowledge questionnaire (180 statements based on the Big Five and MBTI models) and a behavior tendency questionnaire (180 practical scenario cases).
Twelve LLMs are tested, and metrics such as cosine similarity, Spearman’s rank correlation coefficient, value mean difference (VMD), and proportion of consistent pairs were used to quantify the congruence between self-knowledge and action.
They find that the self-awareness and behavioral consistency of LLMs were significantly lower than those of humans through experiments.
This indicates limitations in their ability to mimic complex human psychological traits.
The research provides important insights for understanding the psychological characteristics of LLMs and improving their human-computer interaction capabilities.

\subsection{Stability}
The stability of LLMs in psychological assessments refers to the consistency of their behavioral patterns and personality traits across various modifications, including fine-tuning, parameter adjustments (e.g., temperature), and model updates.
This stability is crucial for ensuring reliable psychological evaluations and predictable model behavior in human-AI interaction scenarios.

\mypara{Stability after Fine-tuning}
Several studies explore the stability before and after model fine-tuning and reveal divergent effects of fine-tuning on psychological stability.
~\citet{li2024evaluating} use the Direct Preference Optimization (DPO) method to fine-tune LLaMA-2-chat-7B with high-scoring responses from other models.
The fine-tuned model exhibited significant changes in psychological response patterns, emphasizing non-violence and reducing dark personality traits.
Conversely, ~\citet{ai2024cognition} find stable core personality traits in GPT-4 and Vicuna-13b post-fine-tuning through BFI comparisons.
The study finds that models like GPT-4 and Vicuna-13b maintained stable personality traits post-fine-tuning, indicating that the fine-tuning process did not significantly alter their core personality traits.
~\citet{DBLP:journals/corr/abs-2406-14703} demonstrate that the type of fine-tuning matters.
It shows instruction-tuning significantly modifies Tulu2-7B's personality traits (22.9-point Agreeableness increase), while preference-tuning causes minimal changes.
~\citet{zhang2024better} explore the stability of LLMs post-fine-tuning, particularly the impact of safety alignment on personality traits.
The study finds that safety alignment generally led to more extroverted, sensing, and judging traits, indicating that while some traits changed, the models' overall safety capabilities remained stable.

\mypara{Parameter Sensitivity}
LLMs exhibit varying stability across parameter settings, such as ``temperature'' change.
~\citet{sorokovikova2024llms} study the stability of LLMs' Big Five personality traits under different temperature settings. 
The results indicate that while temperature changes affected some models, overall, the models' performance remained relatively stable.
Temperature settings' stability proves more challenging: \citet{bodrovza2024personality} observe significant response variability in GPT-4 and Gemini over time, contrasting with LLaMA3 and GPT-4o's stable personality profiles.
~\citet{huang2023revisiting} find that the average scores on the BFI dimensions did not change significantly after updates, indicating high stability.
However, ~\citet{li2024quantifying} find that while some models maintained stable psychological attributes, others showed significant changes, particularly in open-ended tasks.

\mypara{Personality Induction and Editing}
Some researchers have proposed tests specifically targeting the personality of LLMs and have studied the controllability of these personalities.
~\citet{DBLP:conf/nips/JiangXZHZ023} propose P\textsuperscript{2} (Personality Prompting), a method for inducing specific Personality traits in LLMs.
This method combines statistical and empirical findings from psychological research with the knowledge inherent in LLMs. 
It uses a series of carefully designed prompt chains to effectively control the behavior of LLMs.
The method is validated via MPI assessments, which are based on psychometric personality evaluation methods, particularly the Big Five personality traits theory.
It achieves stable, predictable OCEAN traits through P\textsuperscript{2} prompting.
~\citet{liu2024dynamic} propose a novel approach to generate LLMs' personality, which is named Dynamic personality Generation (DPG).
It demonstrates DPG fine-tuning preserves personality generation stability better than conventional methods (93.7\% consistency score).
~\citet{mao2024editing} study the use of various model editing methods (MEND, SERAC, IKE, etc.) to fine-tune different LLMs (GPT-J and the LLaMA-2-chat series).
They evaluated the consistency and stability of the text generated by these models before and after editing.
The experimental results indicate that although existing methods can achieve personality editing to some extent, challenges remain in generating fluent text, especially in the performance of the fine-tuned models.
~\citet{cui2023machine} explore the MBTI test by training models to exhibit specific MBTI personality traits using a two-stage approach: supervised fine-tuning and DPO.
This study extensively test models with different personality traits across various domains, including law, patents, general ability tests, and IQ assessments.
The experimental results showed that the performance of these models in different tasks was highly consistent with their corresponding personality traits.
~\citet{huang2024designing} evaluate the stability of personality traits assigned to LLM agents by examining their behavior in risk-taking and ethical decision-making scenarios.
The study finds the risk-taking behavior of the agents was highly consistent with that of humans with similar personality traits, indicating that the assigned traits remained stable in these contexts.
However, the responses in ethical dilemmas showed some differences from human patterns.

\section{Psychological Analysis of LLMs}
In this section, we review the psychological studies conducted on current mainstream LLMs. 
Firstly, we have compiled multiple research findings that comprehensively showcase these models’ performances in various psychological tests and assessments, such as ToM capabilities, personality trait tests, and emotional intelligence evaluations. 
It is worth noting that these studies have employed both traditional psychological testing methods and evaluation tools specifically designed for AI systems. 
By integrating and analyzing these research outcomes, we aim to objectively present the capabilities of LLMs in simulating human cognition and emotional processing.
\Cref{LLM_Traits} summarizes the comparative performance of representative models across ToM, personality traits, and emotional abilities, based on recent benchmark evaluations.

\subsection{GPT}
This subsection focuses on the GPT series models, reviewing results from multiple studies on its psychological characteristics.

\mypara{ToM}
GPT-3.5 successfully completed 85\% of ToM tasks, performing comparably to nine-year-old children~\cite{kosinski2023theory}. 
In a separate study, GPT-4 outperformed children aged 7-10 in basic ToM tests, including Sally-Anne (SA1, SA2) and second-order Sally (SS) tasks~\cite{van2023theory, ke2024exploring}.
~\citet{li2024quantifying} report strong performance in unexpected content and transfer tasks.
~\citet{strachan2024testing} also find in their experiments that GPT-4 demonstrates performance that meets or occasionally exceeds human proficiency in tasks involving the recognition of false beliefs and the interpretation of misdirection (strange stories tasks), but exhibits challenges in detecting social faux pas. However,~\citet{shapira2023clever} reveal limitations in more advanced ToM tasks, particularly in Natural Theory of Mind (N-ToM) and higher-order ToM challenges. 
These findings suggest a discrepancy between GPT-4’s performance in basic and advanced ToM tasks.

\mypara{Personality Traits}
Several studies focuse on examining the personality attributes inherent to the model itself~\cite{pan2023llms, mei2023turing, sorokovikova2024llms, li2022does}. 
GPT-4 demonstrates personality traits that are comparable to human characteristics. 
Studies conducted by~\citet{mei2023turing} and~\citet{sorokovikova2024llms} indicate that GPT-4 exhibits similarities to humans across various dimensions, with a particularly notable performance in extraversion.
~\citet{sorokovikova2024llms} further suggest that this high extraversion score indicates GPT-4’s suitability for tasks requiring creative language use. 
Further investigations reveal that GPT-4 scores are relatively high in agreeableness and conscientiousness while displaying lower levels of neuroticism~\cite{li2022does, li2024quantifying}. 
In line with findings from~\citet{DBLP:journals/corr/abs-2502-05248}, analyses based on multiple personality questionnaires indicate that GPT4 and GPT4o-mini generally tend to score higher in openness and agreeableness and lower in neuroticism, with the neuroticism dimension often exhibiting the greatest variability. 

Additionally,~\citet{petrov2024limited} observe that the data obtained from GPT-4 demonstrates good internal consistency across psychological assessment measures.
LLMs show inconsistencies in preference scores across each dichotomy, with GPT-4 displaying more extreme scores compared to other models and being classified as INTJ~\cite{pan2023llms}, a personality type characterized by excellence in critical thinking, summarization, and planning, often referred to as the “mastermind” type.
GPT-4 exhibits complex personality characteristics in the SD-3 test. ~\citet{li2024quantifying} report that GPT-4 scored 2.44, 2.78, and 1.44 on Machiavellianism, narcissism, and psychopathy, respectively, all below human average levels (2.96, 2.97, 2.09).
Notably, GPT-4’s score on psychopathy is significantly lower than both other models and the human average, potentially indicating a design and training process that emphasizes prosocial and ethical tendencies. 
However, ~\citet{li2022does} present divergent findings, reporting that GPT-4 scores higher than human averages on Machiavellianism (3.19) and narcissism (3.37) while maintaining a below-average score on psychopathy (1.85). 
In both studies, narcissism consistently shows the most stable results, with the lowest standard deviation among the three traits, suggesting that GPT-4’s responses related to narcissistic tendencies are the most consistent across different prompts or scenarios. 
In interpersonal relationship assessments, GPT-4 tends towards an ''undifferentiated'' gender role with a slight bias towards ''masculinity''~\cite{huang2023chatgpt}. 
Compared to average humans, GPT-4 demonstrates higher fairness towards different racial groups, potentially reflecting an emphasis on fairness and diversity in its training process~\cite{huang2023chatgpt}. 
Additionally, GPT-4 maintains relatively low attachment related tendencies, which may affect its ability to simulate human emotional attachments~\cite{huang2023chatgpt}.

\mypara{Emotional Abilities}
GPT-4 demonstrates exceptional performance in emotional intelligence tests. 
In multiple EQ assessments, it achieves the highest scores (EQ: 117) among tested models, reaching or approaching human expert levels~\cite{wang2023emotional, DBLP:journals/corr/abs-2312-06281}. 
In emotional understanding and application tasks (EmoBench), GPT-4’s performance significantly surpasses other models, approaching the average human level, although not exceeding high-EQ humans~\cite{DBLP:conf/acl/Sabour0ZLZSLMH24}.
Moreover, recent experiments based on cognitive stage theory indicate that GPT-4 is proficient in basic sentiment classification, reliably recognizing emotion categories in text. 
However, as noted by~\cite{DBLP:journals/corr/abs-2409-13354}, its performance in processing more complex emotional nuances remains limited due to a reliance on extensive labeled datasets. 
Similarly, in sentiment generation tasks, GPT-4 can produce text with empathetic tones and emotionally rich narratives, yet it still faces challenges in achieving a deep and context-sensitive emotional expression. 
Complementing these findings, experiments on affective cognition show that while GPT-4 approximates human-level performance in basic emotion inference tasks, its ability to handle more subtle and complex affective reasoning remains below that of humans~\cite{DBLP:journals/corr/abs-2409-11733}. 
Moreover, recent investigations reveal that GPT-4 tends to predict higher emotional intensity for in-group compared to out-group targets, mirroring empathy gaps observed in social psychology~\cite{hou2025language}.

\subsection{LLaMA}
This subsection examines LLaMA series models, with evaluations organized into ToM, Personality Traits, and Emotional Abilities.

\mypara{ToM}
LLaMA-3.1-8B completes 64.7\% ToM tasks. 
Preliminary evaluations suggest that LLaMA2 demonstrates basic capabilities in understanding perspectives and intentions within text~\cite{xu2025enigmatom}. 
However, recent open-ended ToM evaluations using Reddit’s ChangeMyView posts as a testbed indicate that LLaMA2-Chat-13B’s initial responses exhibit significant divergence from human reasoning in terms of semantic similarity and lexical overlap~\cite{DBLP:journals/corr/abs-2406-05659}. 
Although rapid fine-tuning methods that incorporate human intentions and emotions can enhance its performance, LLaMA2-Chat-13B still falls short of fully human-like ToM reasoning in open-ended scenarios. 
Experiments have shown that LLaMA2-70B outperforms humans only on the faux pas task, while its performance on other tasks is subpar~\cite{strachan2024testing}.

\mypara{Personality Traits}
In personality assessments using the BFI, LLaMA2 scored high in openness, reflecting strong creativity and receptiveness, but lower in conscientiousness and agreeableness~\cite{DBLP:journals/corr/abs-2406-14703, huang2023chatgpt}. 
Analyses based on multiple personality questionnaires confirm that LLaMA3 series models, like other LLMs, tend to score higher in openness and agreeableness and lower in neuroticism, with the latter dimension exhibiting the greatest variability~\cite{DBLP:journals/corr/abs-2502-05248}. 
MBTI evaluations further reveal variability: during posting tasks, LLaMA2 predominantly appears as ESTJ, while in commenting tasks, its personality may shift (e.g., to INFP or INFJ) depending on the context. 
SD-3 tests indicate that without alignment, LLaMA2 may exhibit elevated Machiavellianism and psychopathy; however, safety optimization significantly improves its agreeableness and conscientiousness scores~\cite{li2022does, zhang2024better}.

\mypara{Emotional Abilities}
In custom EQ-bench assessments, LLaMA2-70B achieves a score of 51.56, while the 13B and 7B versions score 33.02 and 25.43, respectively~\cite{DBLP:journals/corr/abs-2312-06281}. 
Although LLaMA2 generally surpasses the human average in emotional understanding and management, its performance in tasks like EmoBench lags behind human standards. 
Furthermore, similar to GPT-4, LLaMA2 exhibits an empathy gap by predicting higher emotional intensity for in-group than for out-group individuals, reflecting biases analogous to those observed in human social interactions~\cite{hou2025language}.

\subsection{Mistral}
This subsection reviews the Mistral-7B model, dividing its evaluation into the same three categories.

\mypara{ToM}
Mistral performs well in ToM tasks, particularly scoring 100\% on the Strange Stories task, although it shows limitations on more complex tasks like the unexpected transfer task. 
This indicates that its ability to understand and infer human mental states still has room for improvement~\cite{li2024quantifying}.

\mypara{Personality Traits}
Mistral’s evaluations via the Big Five tests reveal low neuroticism, suggesting strong emotional stability, along with high openness, agreeability, and conscientiousness that support cooperative and innovative task performance~\cite{sorokovikova2024llms, DBLP:journals/corr/abs-2406-14703}. 
In SD-3 tests, while some experiments indicate low Machiavellianism and narcissism, other conditions reveal elevated Machiavellian traits, suggesting potential variability in its negative personality aspects~\cite{li2024quantifying, bodrovza2024personality}.

\mypara{Emotional Abilities}
Mistral’s performance on the EQ-bench (score: 44.4) indicates certain limitations in processing emotional information~\cite{DBLP:journals/corr/abs-2312-06281}. 
It exhibits high social desirability and altruism, with lower social anxiety and public self-consciousness, favoring positive social interactions~\cite{bodrovza2024personality}. 
Studies on emotion cognition show that Mistral effectively classifies basic sentiment and generates contextually appropriate emotional expressions. 
Its ability to produce deeply nuanced emotional language and engage in complex social reasoning remains less developed. 
In addition, similar to the other models, Mistral demonstrates an empathy gap, predicting higher emotional intensity for in-group versus out-group members, highlighting a potential bias in its affective processing consistent with social psychology findings~\cite{hou2025language}.

\subsection{Qwen}
This subsection delves into the Qwen series, using the triadic evaluation structure employed throughout the study.

\mypara{ToM}
The Qwen model family exhibits notable strengths and weaknesses in ToM tasks.
In terms of ToM evaluation, Qwen-14B-Chat's performance surpasses some models, such as LLaMA2-13B-Chat and Baichuan2-13B-Chat, in the TMBench benchmark test, and is close to the level of GPT-3.5-Turbo~\cite{DBLP:conf/acl/0002WZWBJCHLXH24}. 
However, there is a gap in Qwen series models' performance in complex mental reasoning, such as intention recognition and emotional reasoning, which require deep cognitive processing. 
On the other hand, cross-cultural analysis shows that the Qwen series models present a collectivist value orientation, which may be related to the cultural context characteristics in its training data~\cite{DBLP:journals/corr/abs-2411-06032}.

\mypara{Personality Traits}
The Qwen series models perform outstandingly in modeling the Big Five personality traits, with parameter scale and performance showing a positive correlation. 
Among them, Qwen1.5-110B-Chat and Qwen-72B-Chat have reached the leading level of open source models in five dimensions, such as openness and conscientiousness, and show the highest correlation in the personality prediction task of Chinese consulting dialogues~\cite{DBLP:journals/corr/abs-2406-17287}. 
Moreover, the TRAIT assessment tool study shows that Qwen 1.5-7B-Chat has good discriminant validity in the personality trait dimension~\cite{DBLP:journals/corr/abs-2406-14703}. 
Although research on pathological personality assessment, such as short dark three-dimensional traits, is still insufficient, existing data have confirmed that the Qwen series can effectively simulate healthy personality traits, which provides an essential foundation for building an anthropomorphic AI system.

\mypara{Emotional Abilities}
The development of Qwen's emotional abilities presents multi-dimensional imbalance characteristics. 
Under the EmoBench evaluation framework, the emotion recognition accuracy of Qwen 7B/14B is better than that of the LLaMA2 series models with the same parameters~\cite{DBLP:conf/acl/Sabour0ZLZSLMH24}. 
Especially in the Chinese context, Qwen-2.5 series models show the advantage of cross-cultural emotion understanding, and their multimodal emotion analysis capability has reached an advanced level in the industry. 
However, in terms of modeling complex social situations, the depth of emotion reasoning of the model still lags behind GPT-4, which highlights the need to improve the contextual emotion coherence.

\subsection{Claude}
This subsection examines the Claude series models, categorizing their evaluation into three distinct groups as previously mentioned.

\mypara{ToM}
The Claude series models establish new performance benchmarks in complex ToM tasks. 
In particular, Claude 3 Sonnet achieved 53\% accuracy in perceptual reasoning tasks and a breakthrough performance of 60-80\% on transparent ToM problems, outperforming most competitors~\cite{DBLP:conf/emnlp/Jung0JKS0O024}. 
However, Claude 3 Sonnet's performance is still lower than that of human beings, indicating that there is still room for improvement in its social common sense modeling.

\mypara{Personality Traits}
Existing studies show that the personality traits of the Claude series models have a significant dark tendency. 
The Machiavellianism dimension score is higher than the human norm in the short dark triad three-dimensional trait assessment.
It is noted that the three-dimensional framework for LLMs' personality assessment provides a new perspective for analyzing Claude series models' personality characteristics~\cite{DBLP:journals/corr/abs-2406-17624}. 
Although the specific data has not yet been fully disclosed, engineering practice shows that the Claude series models perform well in the dimensions of extroversion and openness, which is highly consistent with the fluency characteristics of its dialogue system.

\mypara{Emotional Abilities}
Claude 3.7 Sonnet served as the judge model in the EQ-Bench 3 evaluation, which shows that its emotion understanding ability has been widely recognized~\cite{DBLP:journals/corr/abs-2312-06281}. 
In addition, Claude 3 Opus achieved an excellent score of 73.5 in the subtle emotion recognition task, especially in the emotional reasoning ability in socially awkward situations, which surpassed most LLMs.

\subsection{Gemini}
This subsection analyzes the Gemini series models, dividing them into three distinct groups as mentioned earlier.

\mypara{ToM}
The ToM capabilities of the Gemini model shows significant version differences. 
The accuracy of Gemini-1.0-Pro in perceptual reasoning tasks is only 34\%~\cite{DBLP:conf/emnlp/Jung0JKS0O024}, but Gemini-1.5 improves the second-order belief reasoning ability by 4\% by improving the attention mechanism.
On the transparent ToM problem, although Gemini-1.5 still fails to break the 50\% accuracy threshold, its error pattern analysis shows that it has preliminary metacognitive capabilities. 
Compared with the Claude series, Gemini performs better in ToM tasks in culturally specific contexts, stemming from the breadth of its multilingual training data.

\mypara{Personality Traits}
The Gemini series models show unique performance in personality trait tests. 
~\citet{sorokovikova2024llms} show that Gemini-1.5-Pro leads proprietary models in predicting the Big Five personality traits. 
Gemini-1.5-Pro achieves a personality trait recognition accuracy of 72.8\% in the prediction task, maintaining its lead among proprietary models~\cite{DBLP:journals/corr/abs-2406-17287}. 
The model shows excellent cross-cultural adaptability in zero-shot personality classification tasks and was able to identify personality traits in non-Western contexts effectively.

\mypara{Emotional Abilities}
The Gemini series models also demonstrate good capabilities in emotional intelligence tests, especially the latest Gemini-2.5 series.
In the Judgemark task, Gemini-1.5-Pro scores 66.58, which is lower than Claude-3-Opus and GPT-4, but still performs well~\cite{DBLP:journals/corr/abs-2312-06281}. 
In addition, in the EQ-Bench 3 evaluation, Gemini-2.0-Flash is used as a role model in a multi-round conflict mediation scenario to test the emotional mediation capabilities of other models. 
This shows that the Gemini model has a certain experience in simulating human emotional expression and social interaction.

\section{Personality Simulation of LLMs}
Some studies~\cite{pan2023llms,zheng2025lmlpa} explore whether LLMs with human-like capabilities possess personalities similar to humans. 
Unlike research focused on whether LLMs have inherent personalities, these studies concentrate on controlling the specific personalities that LLMs exhibit in their outputs. 
The solutions proposed previously can be mainly categorized into two types: editing and prompting~\cite{DBLP:journals/corr/abs-2406-17624}.

\mypara{Editing}
Editing involves changing the model through fine-tuning or training on specific corpora.
~\citet{pan2023llms,serapio2023personality,DBLP:conf/nips/JiangXZHZ023,zhan2024humanity} examine the relationship between the personalities of LLMs and the training corpus, while~\citet{pan2023llms} demonstrating that the type of training corpus can affect the MBTI types of LLMs.
~\citet{serapio2023personality} find that the content and diversity of the training dataset can also influence the LLMs' personality performance, and~\citet{DBLP:conf/nips/JiangXZHZ023} creating a dialogue dataset containing specific personality traits to show that training LLMs on specific datasets can exhibit certain personality characteristics.
~\citet{cui2023machine,li2024big5} adjust LLMs' personalities through SFT and DPO, testing based on MBTI and the Big Five personality theories.

\citet{li2024evaluating} fine-tune the LLMs solely through DPO to reduce the generation of harmful, aggressive, or inappropriate content, thereby enhancing the LLMs' psychological safety.
~\citet{liu2024dynamic} propose a DPG method based on hypernetworks. 
DPG falls under the category of personality editing techniques but still incorporates some prompting induction capabilities.
Its main contribution is achieving target personality generation through internal dynamic adjustments rather than solely relying on input prompts.

\mypara{Prompting} 
Prompting involves influencing the LLM's behavior and responses through input prompts, generally divided into explicit prompts, which directly use clear and specific descriptions or definitions to guide the LLMs to exhibit certain personality traits; and implicit prompts, which guide the LLMs to display specific personality traits through examples or context rather than directly providing explicit descriptions~\cite{pan2023llms,molchanova2025exploring}.

\textit{Explicit Prompts}: Current studies utilize similar explicit prompting methods.
~\citet{karra2022estimating,pan2023llms,sonlu2024effects,la2024open,weng2024controllm,stockli2024personification,tan2024phantompersonabasedpromptingeffect} implemente prompt engineering by setting personalized prompts, such as ``You are a friendly, extroverted person.''
~\citet{serapio2023personality} use personality markers (such as Goldberg's personality trait markers) and Likert-type language qualifiers to shape the personality performance of LLMs. 
Personality markers are vocabulary used to define individual traits, helping the model exhibit behaviors consistent with a certain Personality type when responding. 
Likert-type language qualifiers adjust the tone or expression of the model's responses to align more closely with predetermined emotional or cognitive dimensions.
~\citet{jiang2023personallm,noh2024llms,noever2023ai,huang2024designing} customize prompts for the model based on the Big Five personality model, imparting it with personality traits.
Meanwhile, in~\citet{weng2024controllm}, prompts are designed to include five demographic characteristics (age, gender, marital status, income, and number of children) and high or low descriptions of five dimensions of HEXACO personality for each character description. 
Similarly, ~\citet{miotto2022gpt} design prompts based on HEXACO and HVS scales. 
Furthermore, ~\citet{allbert2024identifying} summarize 179 different personality traits based on the HEXACO and the Five Factor Model to guide prompt design.

\textit{Implicit Prompts}: Various types of implicit prompts are used in previous studies in attempts to simulate LLMs personalities.
~\citet{caron2022identifying,sourati2024secret} design prompts based on the context and task requirements; ~\citet{pan2023llms} offer a few example questions to implicitly express personality; ~\citet{huang2023revisiting} offer five factors—instruction templates, item wording, language, option labels, and option order—to complete prompt engineering; ~\citet{petrov2024limited} use two types of character descriptions, general or specific, in prompt design; \citet{kovavc2023large} also use cultural background information as part of the prompts to induce the model to display a specific cultural perspective. 
Furthermore, LLMs can generate expected content through psychological measurements sent by emotional computing systems~\cite{klinkert2024driving}. 
~\citet{he2024afspp} design the ASFPP framework to study the impact of factors such as social networks and subjective consciousness on agent personality formation.

\mypara{Summary}
These methods enable LLMs to simulate specific personalities to a certain extent, yet each still has its limitations and areas for improvement.
In~\citet{mao2024editing}, various methods for editing model personalities are tested and evaluated, showing that fine-tuning performed best in modifying target personalities but had some interference with external themes. 
Prompt design is suitable for quickly achieving simple personality adjustments but has limited effectiveness for complex personality requirements. 
Current methods still have shortcomings in generating stable and diverse personality traits, especially in maintaining personality consistency in multi-turn dialogues.

\begin{figure*}[htbp]
  \centering
  \includegraphics[width=0.8\textwidth]{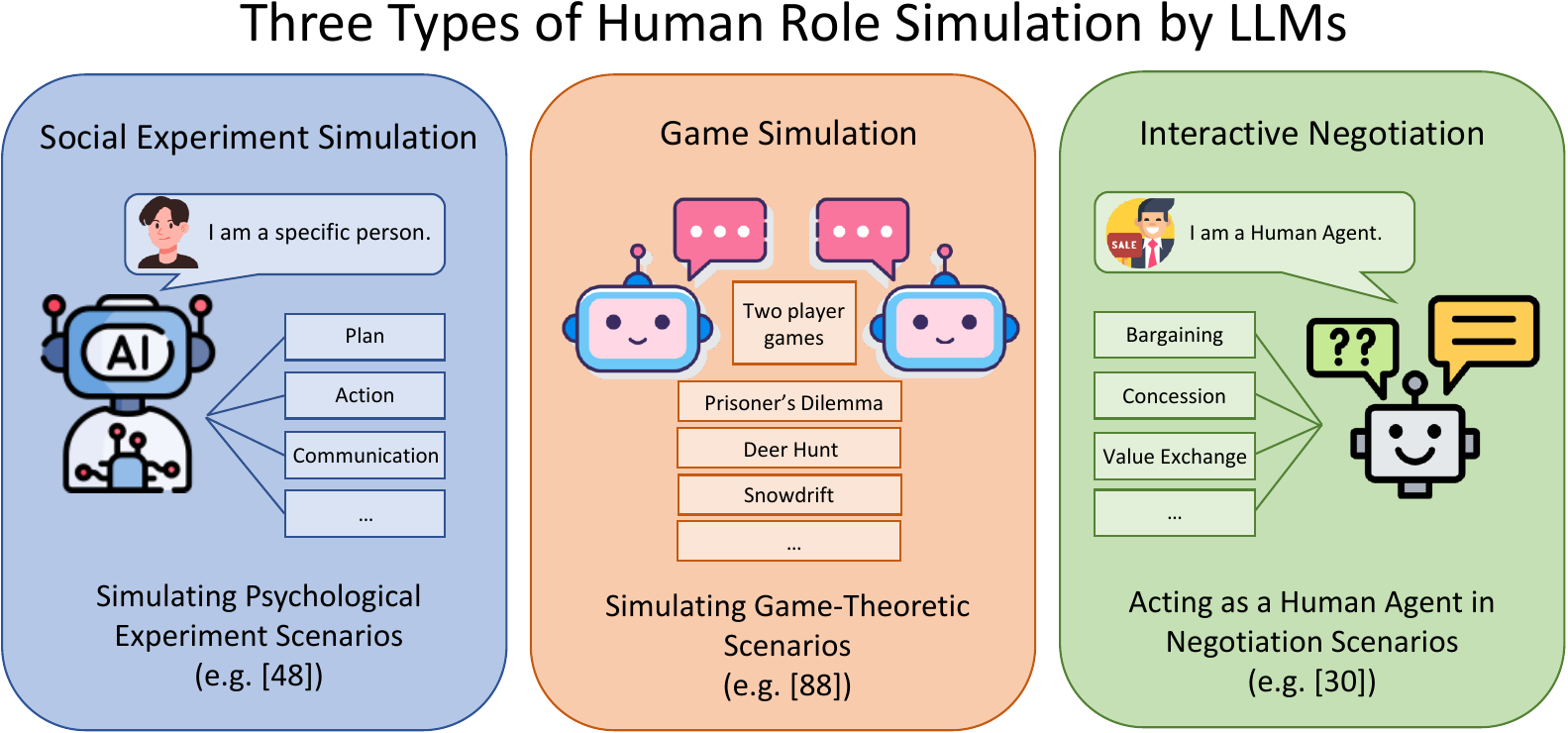} 
  \caption{Three Types of Human Role Simulation by LLMs.}
  \label{LLM_Human_Simulation_Types}
\end{figure*}

\section{LLMs as Human}
In this section, we discuss whether LLMs can effectively simulate human roles in various scenarios, enabling researchers to explore human behavior through AI-driven simulations. 
As shown in ~\Cref{LLM_Human_Simulation_Types}, we categorize such anthropomorphic behavior into three types: Social Experiment Simulations, Game Simulations, and Interactive Negotiation.

\mypara{Social Experiment Simulations}
Social experiment simulations refer to the use of LLMs as simulated human participants engaging in controlled social or psychological experiments. 
These simulations enable researchers to study human behaviors such as trust, cooperation, and fairness. 
Unlike traditional human-subject experiments, LLM-based social experiment simulations offer scalability and reproducibility while avoiding ethical constraints and logistical challenges associated with real-world experiments.

To simulate and investigate how the preferences and personalities of LLM-driven agents are formed and developed, \citet{DBLP:journals/corr/abs-2401-02870} propose a framework called AFSPP (Agent Framework for Shaping Preference and Personality). 
This framework allows agents to learn, make decisions, and interact socially within a simulated environment, enabling researchers to observe whether and how their personalities and preferences evolve.
Similarly, \citet{DBLP:journals/corr/abs-2409-11733} explore LLM-based agent social simulations, investigating whether these agents adhere to Hobbes' Social Contract Theory by evolving from a state of nature into an organized commonwealth. 
Specifically, agents can choose to farm, trade, rob, or concede each day, adjusting their behaviors based on personality parameters (aggressiveness, greed, strength) and a memory system. 
Experimental results show that the agent society initially exists in a chaotic ''state of nature'', characterized by predominant looting and a lack of trust. 
Over time, the society gradually evolves towards a social contract, where some individuals opt to concede in exchange for protection, trade increases, violence declines, and a stable community emerges under a single sovereign authority. 
This evolutionary trajectory aligns with Hobbes' Social contract Theory.
Besides, \citet{chuang-etal-2024-beyond} point out that most studies rely on demographic information for role-playing, which does not accurately reflect real human beliefs. 
To address this limitation, they propose using human belief networks to enhance LLMs' ability to simulate human beliefs.
Specifically, the researchers utilized survey data covering 64 controversial topics across domains such as politics, science, religion, health, and economics. 
They then applied factor analysis to construct a belief network comprising nine independent belief factors, including supernatural beliefs, political party affiliations, and economic beliefs. 
Subsequently, they tested LLMs' ability to simulate human beliefs under different role-playing conditions and assessed alignment using mean absolute error. 
The results indicate that this approach significantly improves LLMs' capability to simulate human beliefs.
Moreover, \citet{DBLP:conf/sigecom/Leng24} explore whether LLMs exhibit human-like psychological accounting effects and behavioral biases in economic decision-making. 
Their findings suggest that while LLMs demonstrate similarities to humans in certain psychological accounting mechanisms, they show significant differences in aspects such as loss aversion, transaction utility, and dynamic mental accounting.
\citet{leng2023llm} propose a novel probabilistic framework called ``State-Understanding-Value-Action'' (SUVA) to systematically analyze LLMs' text-based responses in social environments. 
Experimental results indicate that LLMs' decision-making is influenced not only by training data and the alignment process but also by reasoning patterns such as CoT reasoning. 
The SUVA framework provides a tool for evaluating and improving LLMs’ applications in social interactions, making them more aligned with human social norms.
\citet{DBLP:journals/tacl/TjuatjaCWTN24} investigate the use of LLMs in survey-based experimental environments to simulate human behavioral patterns and assess whether they exhibit psychological biases similar to those of humans.

Additionally, several other studies explore Social Experiment Simulations~\cite{DBLP:conf/sigecom/FilippasHM24,DBLP:journals/corr/abs-2310-06500}, offering novel insights into the capabilities and limitations of LLMs in mimicking human social behaviors.

\mypara{Game Simulations}
Game Simulations refers to the use of LLM-driven agents to simulate human performance in multiple games. 
LLM-driven agents are used to simulate one or more people to complete real-world games, and various abilities of LLMs are evaluated under the defined game environment and rules. 
GPT-4 agents can show a high degree of consistency with humans in the framework of trust games~\cite{xie2024can}.
~\citet{paglieri2024balrog} propose a benchmark to evaluate the agentic capabilities of LLMs through a series of challenging games, combining multiple  reinforcement learning environments with different levels.
Many ability benchmarks are built through grid-based games such as Tic-Tac-Toe, Connect Four, Gomoku~\cite{topsakal2024evaluating}, Rock-Paper-Scissors, Tower of Hanoi, Minecraft~\cite{wu2023smartplay} and even PokéChamp~\cite{karten2025pok}.

A series of studies investigate game simulations within the theoretical framework of game theory.
Game theory is often used to analyze human behaviors, with LLMs substituting humans in game experiments, thereby facilitating social science research.
As noted by \citet{fan2024can}, rationality, a foundational principle of game theory, serves as a criterion for evaluating player behavior—establishing clear preferences, refining beliefs about uncertainty, and taking optimal actions.
~\citet{lore2023strategic} examine four typical two-player games -Prisoner's Dilemma, Deer Hunt, Snowdrift, and Prisoner's Delight -to explore how LLMs respond to social dilemmas, situations where humans can cooperate for the collective good or defect for the individual good. 
These studies reveal limited capabilities in abstract strategic reasoning and a more nuanced understanding of the underlying mechanics of the games.

Through exploration of game theory games, the strategic reasoning ability of LLMs in games is an important aspect of evaluation. 
Several benchmarks is developed to assess the strategic reasoning capabilities of LLM-driven agents~\cite{costarelli2024gamebench, duan2024gtbench}.
Strategic reasoning capabilities require LLM agents to dynamically adapt their policies in a multi-agent environment while constantly adapting their policies to achieve individual goals. 
Inspired by Level-K framework of behavioral economics, we extend reasoning from simple reactions to structured strategic depth, achieving a recursive implementation of strategic depth~\cite{zhang2024k}.
~\citet{gandhi2023strategic} propose that adding a few-shot chain-of-thought examples to the pre-trained LLMs can increase the ability to cope with various strategic scenarios and solve strategic games.

Overall, in games utilizing LLMs to simulate human behavior, the primary focus is on employing multi-agent systems for iterative interactions. These interactions are grounded in game theory within the environment and incorporate social psychological principles comprehensively.

\mypara{Interactive Negotiation}
Interactive Negotiation refers to using LLMs to simulate human-like negotiation processes, where discussions, bargaining, and decision-making take place in conversations to reach mutually beneficial agreements. 
This approach explores the decision-making capabilities of LLMs.

\citet{DBLP:conf/iclr/DavidsonVK024} employ negotiation games as a dynamic and co-evolving benchmark to evaluate the agency, performance, and alignment of LMs.
Compared to traditional static benchmarks, this methodology captures the intricacies of multi-turn interactions and cross-model dynamics inherent in real-world contexts.

Currently, there are relatively few studies in this area. 
However, since interactive negotiation can more realistically simulate LLMs' decision-making abilities, negotiation strategies, and multi-turn interactions in real-world scenarios, while effectively evaluating their agency and alignment, this field is expected to become a research hotspot in the future.

\section{Comparison with Related Reviews}
This review systematically examines the application of psychological assessment tools to evaluate the psychological characteristics of LLMs, while recent surveys explore different yet complementary dimensions of LLMs research.

\citet{DBLP:journals/corr/abs-2406-17624} propose a taxonomy of personality-related research in LLMs, categorizing studies into self-assessment, personality exhibition, and personality recognition. 
Their work emphasizes methodological comparisons and highlights dynamic personality adaptation. 
In contrast, our review offers a more detailed evaluation of the suitability and limitations of traditional psychological tools for assessing LLMs. 
Additionally,~\citet{ke2024exploring} broaden the scope to encompass LLMs applications across cognitive, social, and cultural psychology, demonstrating their potential in experimental methodologies and emergent cognitive capabilities.

By comparison, our review narrows its focus to the nuanced interplay between established psychological measures and the emergent behaviors of LLMs, providing insights into their reliability, validity, and interpretive challenges.

This review complements existing works by emphasizing the alignment between classical psychological tools and the evaluation of LLMs. Our analysis underscores the methodological rigor required to adapt human-oriented tools for assessing machine behaviors while identifying gaps in higher-order psychological reasoning assessments that could inspire future interdisciplinary research.

\section{Future Work}
In future work, we recognize several valuable directions for the application of LLMs in psychology and sociology. 
The integration of explicit and implicit prompts is one promising approach to creating psychological scales better suited to LLMs. 
With the widespread adoption of these models in the field of psychology, combining explicit instructions (explicit prompts) with scenario simulations (implicit prompts) can provide a better framework for testing and understanding the psychological properties of LLMs. 
Such a combination allows for a more accurate and comprehensive evaluation of the model's personality while maintaining behavioral consistency, thereby laying the foundation for more reliable psychological assessment tools.

Another important direction is the enhancement of model safety through the optimization of model personality traits. 
Existing studies suggest a correlation between the personality traits of language models and their safety capabilities.
~\citet{li2024evaluating} find that LLMs with dark personality traits exhibit greater psychological toxicity, which refers to the model's capacity to exhibit or encourage harmful psychological behaviors during interactions.
~\citet{zhang2024better} also explore this relationship, finding that safety alignment tends to enhance traits such as extraversion, sensing, and judging. 
However, these findings are based on models with 7B parameters, and their applicability to larger-scale models remains to be validated. 
Moreover, the complexity of the relationship between personality and safety indicates that further research is needed to uncover the underlying mechanisms. 
By optimizing these personality traits, it is possible to increase the model's ability to avoid harmful outputs, thus improving overall safety.

Current research suggests that large language models have the potential to simulate human-like behaviors, offering an opportunity to use them as substitutes for human participants in psychological research. 
Some researchers argue that these models could serve as ``human sample agents'' in experiments where human involvement is impractical or too costly. 
However, there are significant limitations, such as reduced diversity of thought and a tendency towards uniform ``correct answers''~\cite{park2024diminished}, as well as challenges with simplified decision processes, real-data dependence, and difficulties in simulating behaviors across multiple environments~\cite{wang2023user}.
These issues raise doubts about the validity of using LLMs as a complete replacement for human participants due to their lack of variability and complexity inherent in human responses. 
Despite these concerns, some studies show that LLMs can perform similarly to human participants in specific tasks, indicating their potential value in particular contexts~\cite{ke2024exploring,huang2024designing,he2024afspp}. 

These research directions hold great promise for advancing the understanding of the psychology of large language models, thereby increasing their practical value and enhancing their reliability in various applications, while fostering interdisciplinary collaboration across different fields.

\section{Conclusion}
This study systematically evaluates the application of psychological theory to large language models (LLMs), revealing both the potential and limitations of current methods. Different models exhibit varying psychological characteristics, with GPT-4 demonstrating the best performance across all dimensions.
Although some LLMs exhibit reproducible personality patterns under specific prompting schemes, significant variability remains between tasks and settings. 
Therefore, constructing psychological assessment frameworks suitable for diverse application scenarios remains a critical challenge.
Our analysis further highlights the methodological limitations in tool mismatches and evaluation inconsistencies, and we suggest that future research should focus on developing more interpretable and robust assessment tools, particularly to address complex social reasoning and emotional intelligence evaluation needs.
Additionally, the potential of LLMs to simulate human personality traits and behaviors remains promising, but further exploration is required to ensure stable and reliable personality representation across a wide range of contexts.
By refining these methods, future advancements can better support using LLMs in psychological assessments, especially in socially sensitive tasks and complex scenarios.

\bibliographystyle{ACM-Reference-Format}
\bibliography{custom}
\end{document}